\documentclass[nofootinbib,aps,11pt]{revtex4-1}
\usepackage{graphicx}
\usepackage{hyperref}
\usepackage{cancel}
\usepackage{amssymb}
\usepackage{textcomp}
\usepackage{amsmath}
\usepackage{bm}
\usepackage{times}
\usepackage{epsfig}
\usepackage{color}
\usepackage{slashed}
\usepackage{multirow}
\usepackage{makecell}
\usepackage{ulem}
\newcommand{\BR}{{\rm BR}}
\newcommand{\hc}{{\rm H.c.}}
\newcommand{\eV}{{\rm eV}}
\newcommand{\MeV}{{\rm MeV}}
\newcommand{\GeV}{{\rm GeV}}
\newcommand{\TeV}{{\rm TeV}}
\newcommand{\fb}{{\rm fb}}

\begin{document}
\title{\LARGE Hunting for Heavy Majorana Neutrinos with Lepton Number Violating Signatures at LHC}
\bigskip
\author{Chao Guo$^{1}$}
\email{chaog@mail.nankai.edu.cn}
\author{Shu-Yuan Guo$^{1}$}
\email{shyuanguo@mail.nankai.edu.cn}
\author{Zhi-Long Han$^{1}$}
\email{hanzhilong@mail.nankai.edu.cn}
\author{Bin Li$^{1}$}
\email{libinae@mail.nankai.edu.com}
\author{Yi Liao~$^{1,2,3}$}
\email{liaoy@nankai.edu.cn}

\affiliation{
$^1$~School of Physics, Nankai University, Tianjin 300071, China
\\
$^2$ CAS Key Laboratory of Theoretical Physics, Institute of Theoretical Physics,
Chinese Academy of Sciences, Beijing 100190, China
\\
$^3$ Center for High Energy Physics, Peking University, Beijing 100871, China
}
\date{\today}

\begin{abstract}

The neutrinophilic two-Higgs-doublet model ($\nu$2HDM) provides a natural way to generate tiny neutrino mass from interactions with the new doublet scalar $\Phi_\nu$ ($H^\pm,~H,~A$) and singlet neutrinos $N_R$ of TeV scale. In this paper, we perform detailed simulations for the lepton number violating (LNV) signatures at LHC arising from cascade decays of the new scalars and neutrinos with the mass order $m_{N_R}<m_{\Phi_\nu}$. Under constraints from lepton flavor violating processes and direct collider searches, their decay properties are explored and lead to three types of LNV signatures: $2\ell^\pm 4j+\cancel{E}_T$, $3\ell^\pm 4j+\cancel{E}_T$, and $3\ell^\pm\ell^\mp 4j$. We find that the same-sign trilepton signature $3\ell^\pm4j+\cancel{E}_T$ is quite unique and is the most promising discovery channel at the high-luminosity LHC. Our analysis also yields the $95\%$ C.L. exclusion limits in the plane of the $\Phi_\nu$ and $N_R$ masses at 13 (14) TeV LHC with an integrated luminosity of $100~(3000)~\fb^{-1}$.

\end{abstract}

\maketitle

\section{Introduction}

Neutrino mass and mixing provides robust evidence for physics beyond the standard model (SM). Regarding SM as a low energy effective field theory, the tiny neutrino mass can be incorporated by the unique dimension-five Weinberg operator~\cite{Weinberg:1979sa}. The high energy scale realization of this operator remains however to be a theoretical puzzle. If it is realized at the tree level, there are three possibilities to do so~\cite{Ma:1998dn}, which correspond exactly to the canonical type I~\cite{type1}, type II~\cite{type2}, and type III seesaw~\cite{type3} respectively. The new particles introduced in these seesaws are typically so heavy that they are even beyond the reach of high energy colliders such as the Large Hadron Collider (LHC). To achieve tiny neutrino mass at relatively lower scales, more sophisticated mechanisms have been proposed, such as inverse~\cite{inverse} and linear seesaws~\cite{linear}, pushing the effective neutrino mass operator to even higher dimensions~\cite{highD}, or attributing it to a purely radiative effect~\cite{radiative,1loop,2loop,3loop}; see, e.g., Ref.~\cite{Boucenna:2014zba} for reviews. In these generalized mechanisms, new particles could be at a TeV scale, making their detection in principle possible at LHC.

The demand for very heavy Majorana neutrinos in the type I seesaw arises partly from the difficulty to arrange for a naturally small Yukawa coupling between the light and heavy neutrinos which yields a Dirac mass for neutrinos. To relax the tension, a second Higgs doublet $\Phi_\nu$ is introduced in the so-called neutrinophilic two-Higgs-doublet model ($\nu$2HDM)~\cite{Ma:2000cc,Ma:2002pf,Wang:2006jy,Gabriel:2006ns,Davidson:2009ha,
Bandyopadhyay:2009xa,Marshall:2009bk,Haba:2011nb,Haba:2012ai,Maitra:2014qea,
Chakdar:2014ifa,Seto:2015rma}. By assigning a lepton number to $\Phi_\nu$ but not to the Majorana neutrinos $N_R$, the Yukawa coupling between them and the SM lepton doublet $L$, $\overline{L}\widetilde{\Phi}_\nu N_{R}$, is allowed by the lepton number $U(1)_L$ symmetry, while the SM Higgs doublet $\Phi$ is forbidden to couple to $L$ and $N_R$. Assuming further $U(1)_L$ is softly broken by a bilinear term $\Phi^\dagger\Phi_\nu$ in the scalar potential, $\Phi_\nu$ develops a naturally small vacuum expectation value (VEV) out of that of $\Phi$, so that a tiny neutrino mass becomes possible from a small Dirac mass without requiring a terribly heavy Majorana neutrino. Additional benefits such as a dark matter candidate~\cite{JosseMichaux:2011ba,Mitropoulos:2013fla,Choi:2014tga,
Adulpravitchai:2015mna,Baek:2016wml} and leptogenesis~\cite{JosseMichaux:2011ba,Haba:2011ra,
Haba:2011yc,Chao:2012pt,Haba:2014ita,Clarke:2015hta} have also been actively explored in the framework of $\nu$2HDM.

The above discussion indicates that $\nu$2HDM can be considered as a type-I-like seesaw, but with relative larger Yukawa couplings~\cite{Haba:2010zi} among the SM leptons and the new scalar and heavy Majorana neutrinos. The new particles can now be naturally at a TeV scale and could be within the reach of LHC. The usual lepton number violating (LNV) signature for heavy neutrinos at LHC, $pp\to \ell^\pm N_R \to \ell^\pm\ell^\pm jj$, has been well studied~\cite{Datta:1993nm,Almeida:2000pz,Panella:2001wq,Han:2006ip,delAguila:2008cj,Atre:2009rg,
Dev:2013wba,Alva:2014gxa,Deppisch:2015qwa,Banerjee:2015gca,Ng:2015hba,Leonardi:2015qna,Das:2016hof,
Chakrabortty:2012pp,Antusch:2016ejd}, albeit with significant fine-tuning between tiny neutrino masses and detectable heavy-light neutrino mixing in type I seesaw (see, e.g., Refs.~\cite{Xing:2009in,He:2009ua} for discussions on fine-tuning issues). In $\nu$2HDM, the direct production of $\ell^\pm N_R$ is still suppressed heavily by the small mixing, but now $N_R$ could also arise as a decay product of other new particles that can be copiously produced at LHC~\cite{Keung:1983uu,Huitu:2008gf,Basso:2008iv,Cerdeno:2009dv,Basso:2011hn,
Das:2012ii,Han:2012vk,Gluza:2015goa,Gago:2015vma,Kang:2015uoc,Lindner:2016lxq,
Mitra:2016kov,Accomando:2016rpc}.
This is indeed the case when $N_R$ is lighter than $\Phi_\nu$ which can be pair or associated produced via the Drell-Yan process followed by its decay to $N_R$ via the Yukawa coupling. The further decay of $N_R$, $N_R\to \ell^\pm W^\mp$, then triggers various LNV signatures at LHC~\cite{Wang:2016vfj}. The purpose of this work is to investigate these signatures by detailed simulations.

Our paper is organized as follows. In Sec.~\ref{MD}, we introduce $\nu$2HDM and consider constraints from lepton flavor violation (LFV) and direct collider searches. Then in Sec.~\ref{DP}, we study the decay properties of neutrinophilic scalars and heavy neutrinos. The detailed simulation of various LNV signatures at LHC is performed in Sec.~\ref{SG}. Finally, our conclusions are presented in the last Sec.~\ref{CL}.

\section{Model and Constraints}\label{MD}

\subsection{The Model}

The $\nu$2HDM was previously suggested in Ref.~\cite{Ma:2000cc}. It introduces a new scalar doublet $\Phi_\nu$ that has the same quantum numbers as the SM Higgs doublet $\Phi$, and three right-handed singlet neutrinos $N_{R}$. A global $U(1)_L$ symmetry is then imposed, under which $\Phi_\nu$ has lepton number $L= -1$ and $N_{R}$ has null lepton number. This then distinguishes between $\Phi_\nu$ and $\Phi$: while $\Phi_\nu$ couples to $N_R$, $\Phi$ couples only to SM fermions, thus avoiding flavor changing neutral currents at tree level in the Yukawa sector. When a soft $U(1)_L$ breaking term is introduced in the scalar potential, a small VEV of $\Phi_\nu$ can naturally develop, making $\nu$2HDM generically different from the conventional two-Higgs-doublet models~\cite{Branco:2011iw}.

Denoting the two scalar doublets as
\begin{align}
\Phi=\left(
\begin{array}{c}
\phi^+\\ (v+\phi^{0,r}+i \phi^{0,i})/\sqrt{2}
\end{array}\right),~
\Phi_\nu=\left(
\begin{array}{c}
\phi^+_\nu\\ (v_{\nu}+\phi^{0,r}_{\nu}+i \phi^{0,i}_{\nu})/\sqrt{2}
\end{array}\right),
\end{align}
the scalar potential is
\begin{eqnarray}
V & = & -m_{\Phi}^2 \Phi^\dag \Phi +  m_{\Phi_\nu}^2 \Phi^\dag_\nu \Phi_\nu
       +\frac{1}{2}\lambda_1(\Phi^\dag \Phi)^2
       +\frac{1}{2}\lambda_2(\Phi^\dag_\nu \Phi_\nu)^2
\nonumber
\\
 & & +\lambda_3(\Phi^\dag \Phi)(\Phi^\dag_\nu \Phi_\nu)
 +\lambda_4(\Phi^\dag \Phi_\nu)(\Phi^\dag_\nu \Phi)
 - \left(\mu^2 \Phi^\dag\Phi_\nu +\hc \right).
\end{eqnarray}
Since vanishing of $\mu^2$ would enhance symmetry, it can be considered as naturally small. Assuming $m^2_{\Phi,\Phi_\nu}>0$ and $\lambda$s fulfil various bounded-from-below conditions~\cite{Gunion:2002zf}, $\Phi_\nu$ develops a small VEV out of that of $\Phi$:
\begin{equation}
v\simeq \sqrt{\frac{2 m_{\Phi}^2}{\lambda_1}},~
v_\nu \simeq \frac{\mu^2 v}{m_{\Phi_\nu}^2+(\lambda_3+\lambda_4)v^2/2}.
\end{equation}
To get some feel about the seesaw-like relation between the VEVs $v_\nu$ and $v$, we may assume, for instance, $m_{\Phi_\nu}\sim~500~\GeV$, $\mu^2\sim~10~\GeV^2$ to arrive at $v_\nu\sim~10~\MeV$. Since $\mu^2$ is the only source of $U(1)_L$ breaking, its radiative correction is proportional to itself and only logarithmically sensitive to the cutoff~\cite{Davidson:2009ha}. The hierarchy $v_\nu\ll v$ is thus stable against radiative corrections~\cite{Morozumi:2011zu,Haba:2011fn,Haba:2011yc}.

After electroweak symmetry breaking, the physical scalars are given by
\begin{eqnarray}
H^\pm=\phi^\pm\sin\beta-\phi^\pm_\nu\cos\beta&,~& A=\phi^{0,i}\sin\beta-\phi^{0,i}_\nu\cos\beta,
\\
H=\phi^{0,r}\sin\alpha-\phi^{0,r}_\nu\cos\alpha&,~& h=-\phi^{0,r}\cos\alpha-\phi^{0,r}_\nu\sin\alpha,
\end{eqnarray}
where the mixing angles $\beta$ and $\alpha$ are determined by
\begin{equation}
\label{mix}
\tan\beta=\frac{v_\nu}{v},~\tan2\alpha\simeq2\frac{v_\nu}{v}
\frac{-\mu^2+(\lambda_3+\lambda_4)vv_\nu}{-\mu^2+\lambda_1 vv_\nu},
\end{equation}
and their masses are
\begin{eqnarray}
m_{H^\pm}^2\simeq m_{\Phi_\nu}^2\!+\frac{1}{2}\lambda_3v^2,~
m_A^2\simeq m_H^2\simeq m_{H^\pm}^2\!+\frac{1}{2}\lambda_4v^2,~
m_h^2\simeq 2m_{\Phi}^2=\lambda_1 v^2.
\end{eqnarray}
Since $v_\nu\ll v$ in our consideration here, $\Phi$ is almost identical with the SM Higgs doublet, and $h$ can be regarded as the boson of mass $m_{h}=125~\GeV$ discovered at LHC~\cite{Aad:2012tfa,Chatrchyan:2012xdj,Aad:2015zhl}. For simplicity, we will assume in our numerical analysis a degenerate mass spectrum of $\Phi_\nu$, i.e., $m_{H^\pm}=m_{H}=m_A=m_{\Phi_\nu}$, which can be realized by taking $\lambda_3=\lambda_4=0$.

Under the $U(1)_L$ symmetry, the Yukawa coupling and the Majorana mass terms for $N_R$ are given by
\begin{equation}
\label{yuk}
-\mathcal{L}_N=y\overline{L} \widetilde{\Phi}_\nu N_{R}
+\frac{1}{2}\overline{N_{R}^c}m_{N_R} N_{R}
+ \hc,
\end{equation}
with $\widetilde{\Phi}_\nu=i\sigma_2 \Phi_\nu^*$. Without loss of generality, we assume the charged leptons and $N_R$ have been diagonalized. Note that the same Yukawa coupling can also be obtained by imposing a discrete $\mathbb{Z}_2$ symmetry~\cite{Haba:2011nb}. But the $\mathbb{Z}_2$ scenario is found to be in severe tension with the electroweak precision tests, while the $U(1)_L$ scenario is still viable~\cite{Machado:2015sha}. Analogously to the type I seesaw, the mass matrix for light neutrinos can be derived approximately from Eq.~(\ref{yuk}):
\begin{equation}
\label{eq:mv}
m_\nu = - \frac{1}{2}v_\nu^2 y~ m_{N_R}^{-1} y^T
 = U_{\text{PMNS}}\, \hat{m}_\nu U^T_{\text{PMNS}},
\end{equation}
where $\hat{m}_\nu=\textrm{diag}(m_1,m_2,m_3)$ is the diagonalized neutrino mass matrix and $U_{\text{PMNS}}$ is the usual Pontecorvo-Maki-Nakagawa-Sakata (PMNS) matrix:
\begin{align}
U_{\text{PMNS}}\! =\! \left(
\begin{array}{ccc}
c_{12} c_{13} & s_{12} c_{13} & s_{13} e^{i\delta}\\
-s_{12}c_{23}-c_{12}s_{23}s_{13}e^{-i\delta} & c_{12}c_{23}-s_{12}s_{23}s_{13} e^{-i\delta} & s_{23}c_{13}\\
s_{12}s_{23}-c_{12}c_{23}s_{13}e^{-i\delta} & -c_{12}s_{23}-s_{12}c_{23}s_{13}e^{-i\delta} & c_{23}c_{13}
\end{array}
\right)\!\times\!
\text{diag}(e^{i \varphi_1/2},1,e^{i\varphi_2/2}).
\end{align}
Here the shortcuts $c_{ij}=\cos\theta_{ij}$ and $s_{ij}=\sin\theta_{ij}$ are used, $\delta$ is the Dirac phase and $\varphi_{1,2}$ are the two Majorana phases. In our numerical sampling we will consider neutrino masses in either normal (NH) or inverted hierarchy (IH), and scan randomly the oscillation parameters in the $2\sigma$ ranges of Table I in Ref.~\cite{Forero:2014bxa}.

The Yukawa coupling matrix $y$ in Eq. (\ref{eq:mv}) can be solved in terms of neutrino parameters, $v_\nu$, $m_{N_R}$, and free parameters in the Casas-Ibarra parametrization~\cite{Casas:2001sr,Ibarra:2003up}:
\begin{equation}\label{yexp}
y=\frac{\sqrt{2}}{v_\nu}U_{\text{PMNS}}\hat{m}_\nu^{1/2} R m_{N_R}^{1/2},
\end{equation}
where $R$ is a generalized orthogonal matrix:
\begin{align}
R= \left(
\begin{array}{ccc}
u_{21} & -\omega_{21} & 0\\
\omega_{21} & u_{21} & 0 \\
0 & 0 & 1
\end{array}
\right)
\left(
\begin{array}{ccc}
u_{31} & 0 & -\omega_{31}\\
0 & 1 & 0 \\
\omega_{31} & 0 & u_{31}
\end{array}
\right)
\left(
\begin{array}{ccc}
 1& 0 &0 \\
0 & u_{32} & -\omega_{32} \\
 0& \omega_{32} & u_{32}
\end{array}
\right),
\end{align}
with $u_{ij}=(1-\omega_{ij}^2)^{1/2}$ and $-1\leq\omega_{ij}\leq1$ when $R$ is real. Then the mixing matrix between heavy and light neutrinos can be expressed as~\cite{Perez:2009mu}
\begin{equation}
\label{VlN}
V_{\ell N}=\frac{yv_\nu}{\sqrt{2}}m_{N_R}^{-1}=U_{\text{PMNS}} \hat{m}_\nu^{1/2} R~ m_{N_R}^{-1/2}.
\end{equation}
Differently from the type I seesaw, the small neutrino mass may be attributed to a small $v_\nu$ instead of a large $m_{N_R}$ or a tiny $y$, making heavy Majorana neutrinos possibly within the reach of LHC. For instance, $m_\nu\sim0.01~\eV$ can be obtained with $v_\nu\sim 10~\MeV$, $m_{N_R}\sim 200~\GeV$, and $y\sim 0.006$ whence we have $V_{\ell N}\sim 10^{-7}$.

\subsection{Constraints}\label{constraints}

Now we consider experimental constraints on the $\nu$2HDM. Cosmological considerations have been used to set an upper bound on the sum of neutrino masses, $\sum_i m_i<0.23~\eV$~\cite{Lesgourgues:2012uu,Ade:2013zuv}. The null result searching for neutrinoless double-$\beta$ decays requires the standard effective neutrino mass
$\langle m\rangle_{ee}=|\sum_i (V_{\text{PMNS}})^2_{ei}m_i|$ to be less than $0.061-0.165~\eV$~\cite{Bilenky:2012qi,Auger:2012ar,Gando:2012zm}, while the additional contribution from $\Phi_\nu$ is negligible due to its neutrinophilic nature. It is thus safe to state that the lightest neutrino mass should be less than $0.1~\eV$.

\begin{figure}
\begin{center}
\includegraphics[width=0.45\linewidth]{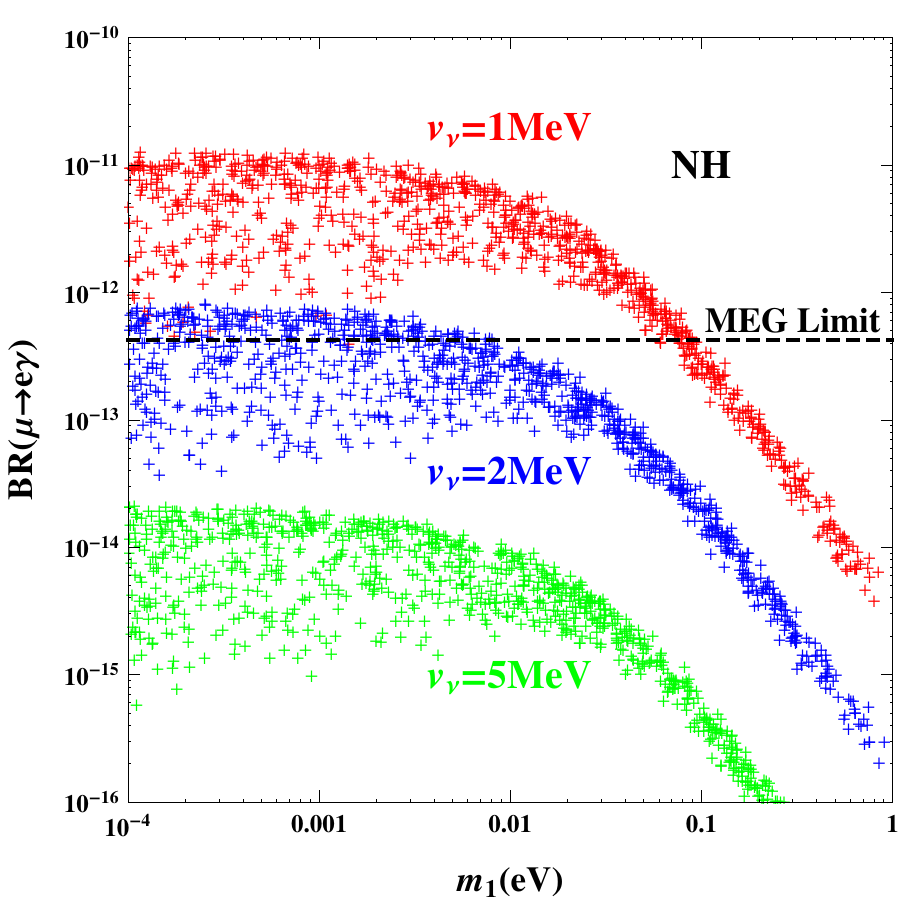}
\includegraphics[width=0.45\linewidth]{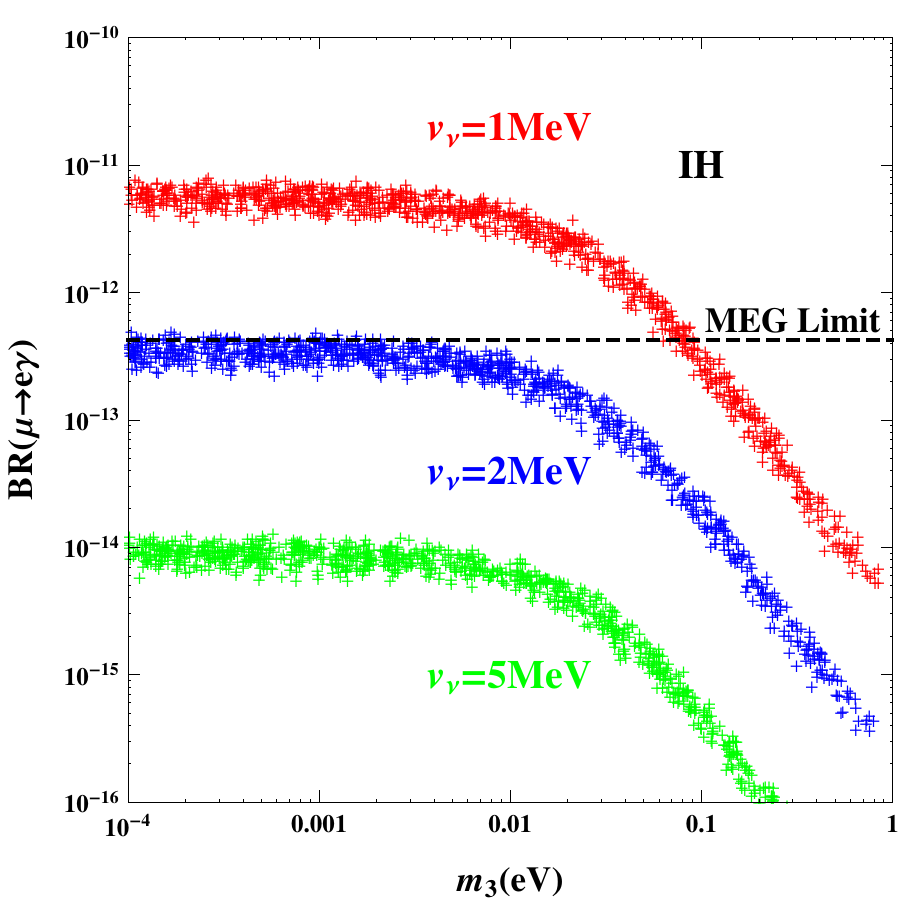}
\end{center}
\caption{$\BR(\mu\to e\gamma)$ as a function of the lightest neutrino mass for normal (left panel) and inverted (right) hierarchy at $(m_{H^+},m_{N_R})=(300,200)~\GeV$.}
\label{LFV}
\end{figure}

The Yukawa coupling $y\overline{L}\widetilde{\Phi}_\nu N_{R}$ with a not too small $y$ can mediate measurable LFV processes~\cite{Ma:2001mr,Bertuzzo:2015ada}. The currently most stringent constraint comes from the MEG experiment on the decay $\mu\to e\gamma$ with $\BR(\mu\to e\gamma)<4.2\times10^{-13}$~\cite{Adam:2013mnn}, which is calculated as \cite{Ma:2001mr,Ding:2014nga}
\begin{equation}
\mbox{BR}(\mu\to e\gamma)=\frac{3\alpha}{64\pi G_F^2}\left|\sum_i \frac{y_{\mu i}y_{e i}^*}{m_{H^+}^2} F\left(\Delta^{N_{Ri}}_{H^+}\right)\right|^2,
\end{equation}
where $\Delta^A_B=m^2_A/m^2_B$ is the ratio of the squared masses and the loop function $F(x)$ is
\begin{equation}
F(x)=\frac{1}{6(1-x)^4}\left(1-6x+3x^2+2x^3-6x^2\ln x\right).
\end{equation}
While nondegenerate $N_R$s are required to generate nondegenerate light neutrinos, we assume approximately in our numerical estimates that they are nearly degenerate to reduce the number of free parameters. The branching ratio then simplifies to
\begin{equation}
\label{BRLFV}
\text{BR}(\mu\to e\gamma)\approx\frac{3\alpha}{16\pi G_F^2} \frac{m_{N_R}^2
|\widetilde{m}_{\mu e}|^2}{m_{H^+}^4 v_\nu^4} \left|F\left(\Delta^{N_{R}}_{H^+}\right)\right|^2,
\end{equation}
with $\widetilde{m}=U_{\text{PMNS}}\hat{m}_\nu U^\dag_{\text{PMNS}}$ and in particular
\begin{equation}
\tilde m_{\mu e}=c_{12}c_{13}c_{23}s_{12}(m_2-m_1)
+c_{13}s_{13}s_{23}e^{-i\delta}[(m_3-m_2)+c_{12}^2(m_2-m_1)].
\end{equation}
Note that the dependence on the free Majorana phases and real matrix $R$ drops out in this approximation, but the dependence on the Dirac phase $\delta$ can be significant for LFV~\cite{Toma:2013zsa} as the mixing angle $\theta_{13}$ is known to be not small.

We show in Fig.~\ref{LFV} $\BR(\mu\to e\gamma)$ as a function of the lightest neutrino mass in either NH or IH and for $(m_{H^+},m_{N_R})=(300,200)~\GeV$. A too small value of $v_\nu\sim 1~\MeV$ is obviously in conflict with the MEG bound. For nearly degenerate light neutrinos with the lightest mass $m_{1/3}\gtrsim 0.01~\eV$, we have $m_j-m_i\approx\Delta m_{ji}^2/(2m_i)$, which explains why $|\tilde m_{\mu e}|$ and thus $\BR(\mu\to e \gamma)$ decrease as $m_{1/3}$ increases. On the other hand, for the lightest mass $m_{1/3}\lesssim 0.01~\eV$, we have $|m_j-m_i|\approx \sqrt{|\Delta m_{ji}^2|}$ with $m_i$ being the lightest mass, so that $\widetilde{m}_{\mu e}$ and thus $\BR(\mu\to e \gamma)$ saturate to a constant. Using Eq.~(\ref{BRLFV}), the upper bound on $\BR(\mu \to e \gamma)$ can be translated into a lower bound on $m_{H^+}v_{\nu}$ for a given $m_{N_R}$:
\begin{equation}
m_{H^+}v_\nu \gtrsim \left[\frac{1\times10^{-13}}{\text{BR}(\mu\to e\gamma)}\left(\frac{m_{N_R}}{100~\GeV}\right)^2\right]^{1/4}\times 600~\GeV\cdot\MeV,
\end{equation}
with $F(m_{N_R}^2/m_{H^+}^2)\sim 0.1$ when $m_{N_R}\sim m_{\Phi_\nu}$. Hence, for $m_{N_R}\sim 200~\GeV$ for instance, the current MEG limit implies that $m_{H^+}v_\nu\gtrsim 600~\GeV\cdot\MeV$, which can also be seen in Fig.~\ref{LFV}. Note that due to the existence of heavy Majorana neutrino $N_R$, the lower bound on $v_\nu$ is about $\sqrt{m_{N_R}/m_\nu}\sim10^6$ times higher than those on the VEV of the Higgs triplet in type II seesaw~\cite{Fukuyama:2009xk} or of the neutrinophilic doublet in the Dirac scenario of $\nu$2HDM \cite{Bertuzzo:2015ada}.

Next we summarize direct collider searches for the new scalar doublet $\Phi_\nu$ and heavy Majorana neutrinos $N_R$. The collider signatures of $\nu$2HDM have been studied in Refs.~\cite{Davidson:2009ha,Davidson:2010sf,Haba:2011nb,Wang:2016vfj}, which concentrated mainly on the charged scalars $H^\pm$. When $m_{N_R}>m_{H^+}$, the dominant decay of $H^+$ would be $H^+\to\ell^+\nu$. The direct searches for signatures such as $\ell^+\ell^-+\cancel{E}_T$ at LHC have excluded the region of $m_{H^+}\lesssim 300~\GeV$~\cite{Aad:2014vma,Khachatryan:2014qwa}. In the opposite case of $m_{N_R}<m_{H^+}$, the dominant decay of $H^+$ would be $H^+\to \ell^+ N_R$ with $N_R$ further decaying into $\ell^\pm W^\mp$, $\nu Z$, and $\nu h$. A brief discussion on this case has been carried out in Refs.~\cite{Haba:2011nb,Wang:2016vfj}, however no searches for such signatures have been performed by experiments. Here we consider the loose LEP bound $m_{H^+}>80~\GeV$~\cite{Abbiendi:2013hk}. Since heavy Majorana neutrinos $N_R$ also exist in the type I seesaw, their searches have been extensively studied via the LNV signature
$pp\to \ell^\pm N_R \to \ell^\pm\ell^\pm jj$~\cite{Keung:1983uu,Han:2006ip,delAguila:2008cj,Atre:2009rg,
Dev:2013wba,Alva:2014gxa,Deppisch:2015qwa,Banerjee:2015gca,Das:2016hof}. For $m_{N_R}<m_W$, LEP excluded  $|V_{\ell N}|^2\gtrsim2\times10^{-5}$~\cite{Abreu:1996pa,Achard:2001qv}, while for heavier $N_R$, LHC could give the most restrictive direct limits, i.e., $|V_{\ell N}|^2<0.017$ at $m_{N_R}=200~\GeV$~\cite{Chatrchyan:2012fla,Aad:2015xaa,Khachatryan:2016olu}.
In $\nu$2HDM, the mixing $V_{\ell N}$ is predicted as $V_{\ell N}=U_{\text{PMNS}} \hat{m}_\nu^{1/2} R~ m_{N_R}^{-1/2}\sim10^{-7}$ for electroweak scale $m_{N_R}$ without fine-tuning, which is far below the current limits.

\section{Decay Properties}\label{DP}

The decay properties of the neutrinophilic scalars $H^\pm$, $H$, $A$ and heavy Majorana neutrinos $N_R$ have been discussed in Refs.~\cite{Ma:2000cc,Haba:2011nb}. For completeness, we first review briefly the scenario with $m_{N_R}>m_{\Phi_\nu}$. Then we concentrate on the opposite case with $m_{N_R}<m_{\Phi_\nu}$, where LNV signatures at LHC can arise.

When $m_{N_R}>m_{\Phi_\nu}$, $N_R$ decays dominantly into $\ell^\pm H^\mp$ and $\nu H/A$. The singlet nature of $N_R$ makes it hardly producible at LHC and thus practically undetectable, but the doublet scalars can be pair or associated produced via the Drell-Yan processes. The scalars may decay as $H^+\to\ell^+\nu$ and $H/A\to\nu\nu$ via the small mixing $V_{\ell N}$ between the heavy and light neutrinos, which would lead to the same dilepton signature $\ell^+\ell^-+\cancel{E}_T$ as in the Dirac type $\nu$2HDM~\cite{Davidson:2009ha,Davidson:2010sf}. But this is not the case for the Majorana type $\nu$2HDM with $U(1)_L$ symmetry when we take into account the constraints from LFV. From Sec. \ref{constraints}, we know that $v_\nu\gtrsim 1~\MeV$ should be satisfied for $m_{\Phi_\nu}$ around the electroweak scale, and then the neutrinophilic scalars decay dominantly as $H^+\to c\bar{b}/t\bar{b}/W^+h$, $H\to b\bar{b}/t\bar{t}/hh$, and $A\to b\bar{b}/t\bar{t}/Zh$ when $v/v_\nu\lesssim10^5$~\cite{Haba:2011nb}. The neutrinophilic scalars in this case are therefore long-lived since both $V_{\ell N}$ and $v_\nu$ are tiny, and the decays of $H^\pm$ could lead to detectable displaced vertices at LHC.

Now we turn to the more interesting scenario $m_{N_R}<m_{\Phi_\nu}$. In this scenario, we can safely neglect the mixing between $\Phi_\nu$ and $\Phi$ with $v_\nu\sim\mathcal{O}(10~\MeV)$~\cite{Haba:2011nb}, so that the dominant decays of the neutrinophilic scalars are $H^+\to \ell^+ N_R$ and $H/A\to \nu N_R$, with heavy Majorana neutrinos $N_R$ decaying further into $\ell^\pm W^\mp$, $\nu Z$, and $\nu h$. These decays are analyzed in the following subsections.

\subsection{Neutrinophilic Scalars}

\begin{figure}
\begin{center}
\includegraphics[width=0.45\linewidth]{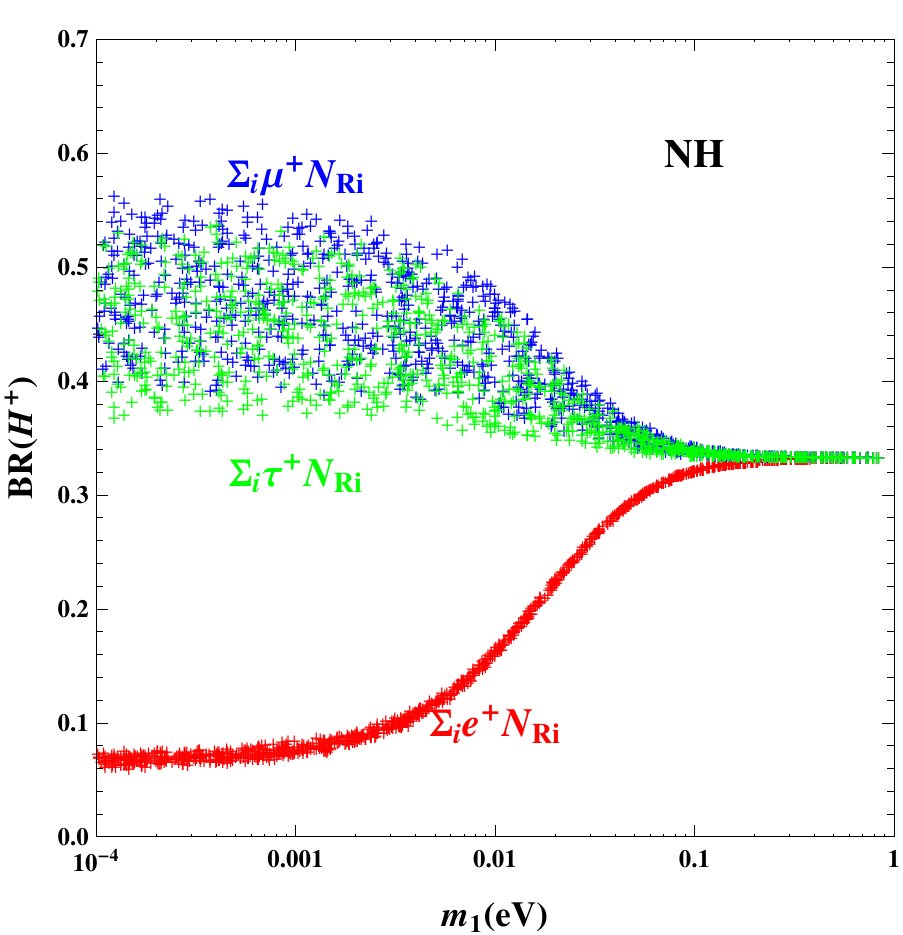}
\includegraphics[width=0.45\linewidth]{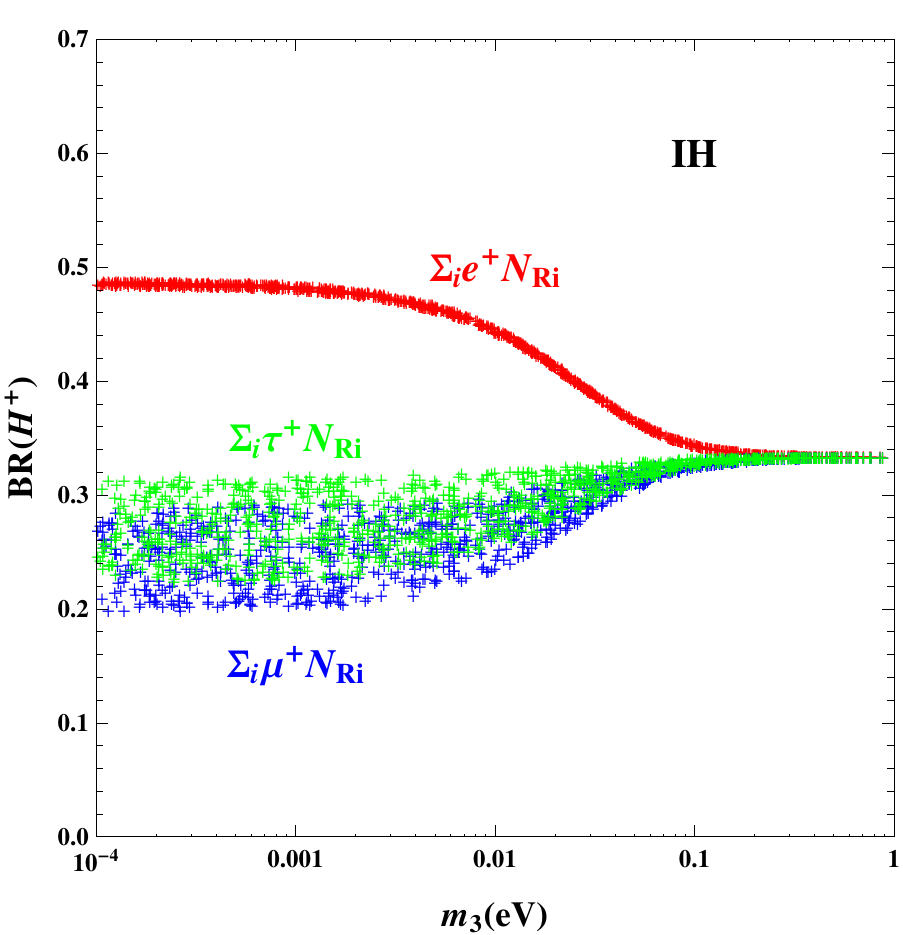}
\end{center}
\caption{Branching ratios of the charged scalar $H^+$ as a function of the lightest neutrino mass for normal (left panel) and inverted (right) hierarchy.}
\label{BRHP}
\end{figure}

In the scenario of $m_{N_R}<m_{\Phi_\nu}$, the neutrinophilic scalars decay into charged leptons or neutrinos and heavy Majorana neutrino $N_R$ via the Yukawa coupling $y$. The partial decay widths are
\begin{eqnarray}
\Gamma(H^+\to \ell^+ N_{Ri}) &=& \frac{|y_{\ell i}|^2}{16\pi} m_{H^+} \left(1-\Delta^{N_{Ri}}_{H^+}\right)^2,
\\
\Gamma(H/A\to \nu_\ell N_{Ri}) &=&\frac{|y_{\ell i}|^2}{16\pi}m_{H/A} \left(1-\Delta^{N_{Ri}}_{H/A}\right)^2.
\end{eqnarray}
The branching ratios of the neutrinophilic scalars are only proportional to $|y_{\ell i}|^2$ and are then determined by the neutrino parameters via Eq.~(\ref{yexp}). As mentioned earlier, we randomly scan the oscillation parameters in their $2\sigma$ ranges of Ref.~\cite{Forero:2014bxa} when evaluating the branching ratios. In Fig.~\ref{BRHP}, we show the scanning results of $\BR(H^+\to \ell^+ N_{Ri})$ as a function of the lightest neutrino mass $m_{1/3}$ in normal/inverted hierarchy by summing over the heavy Majorana neutrinos. We learn that in the nondegenerate neutrino mass region $m_{1/3}\lesssim 0.1~\eV$:
\begin{eqnarray}
\sum_i\text{BR}(H^+\to e^+ N_{Ri})&<&\sum_i\text{BR}(H^+\to \mu^+ N_{Ri})\approx\sum_i\text{BR}(H^+\to \tau^+ N_{Ri})~\text{for NH},\\
\sum_i\text{BR}(H^+\to e^+ N_{Ri})&>&\sum_i\text{BR}(H^+\to \mu^+ N_{Ri})\approx\sum_i\text{BR}(H^+\to \tau^+ N_{Ri})~\text{for IH}.
\end{eqnarray}

\begin{figure}
\begin{center}
\includegraphics[width=0.45\linewidth]{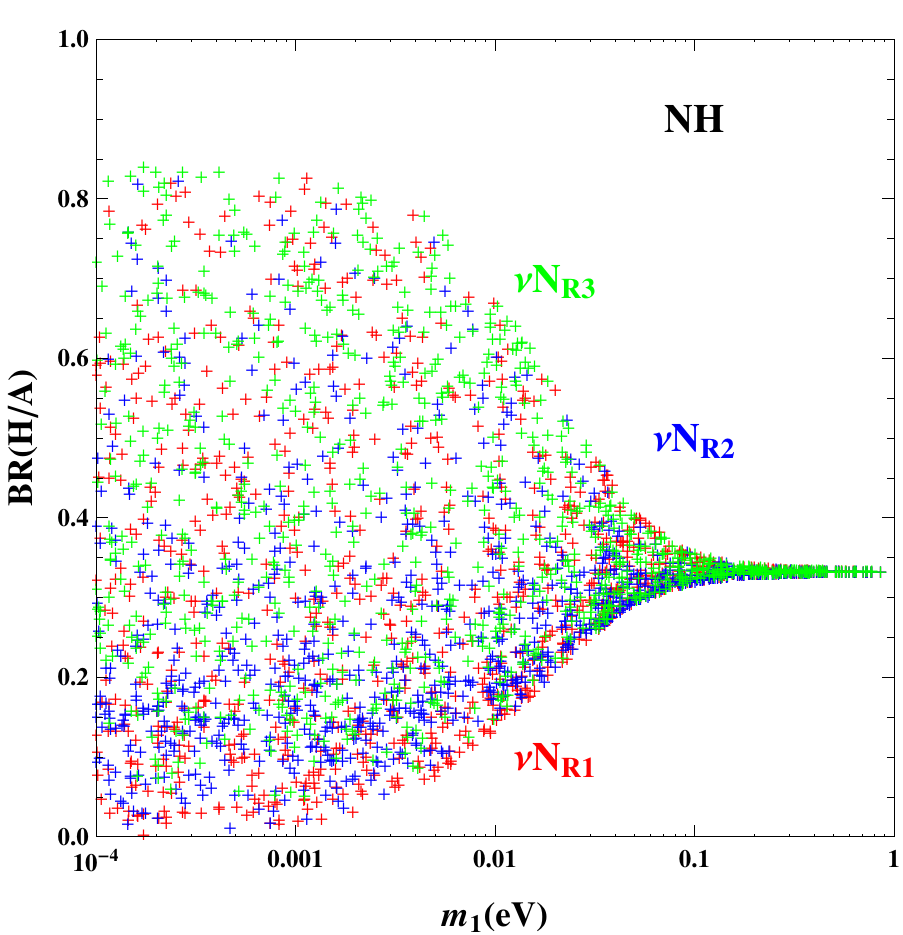}
\includegraphics[width=0.45\linewidth]{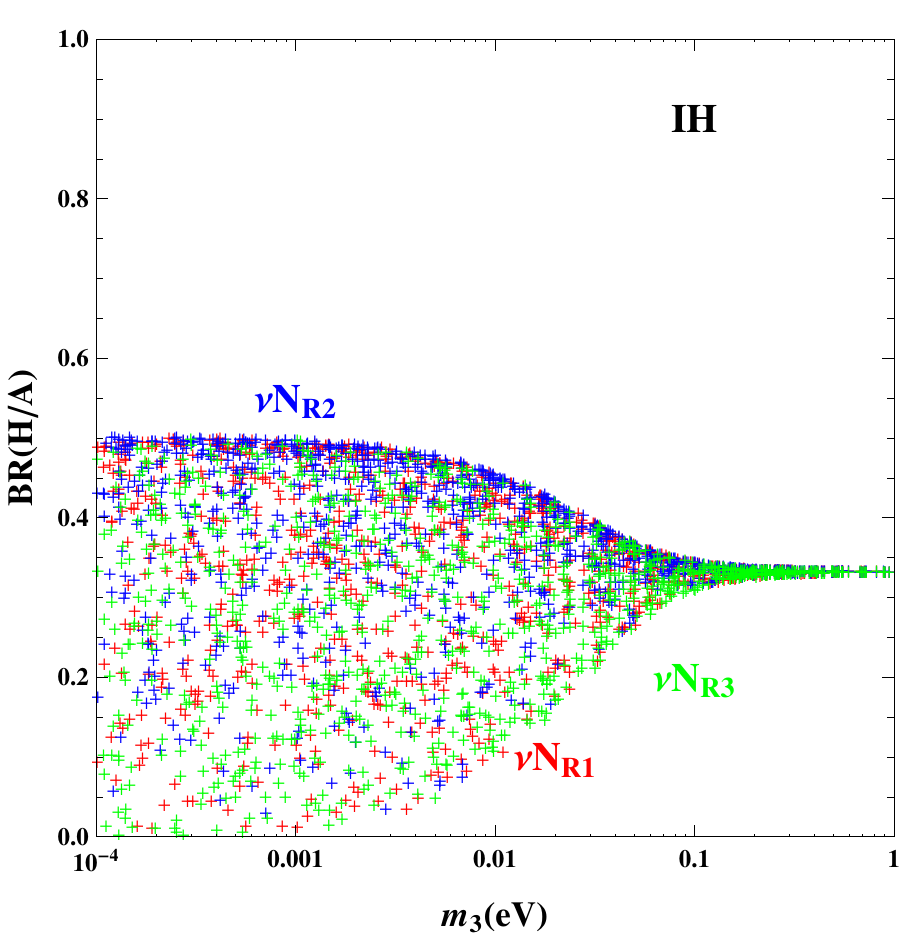}
\end{center}
\caption{Branching ratios of the neutral scalars $H/A$ as a function of the lightest neutrino mass for normal (left panel) and inverted (right) hierarchy.}
\label{BRHA}
\end{figure}

So we expect that the neutrino mass hierarchy might be distinguishable at LHC by the decay products of the charged scalars $H^\pm$. To illustrate this, we use the best-fit values of the neutrino oscillation parameters in Ref.~\cite{Forero:2014bxa} with the lightest neutrino mass $m_{1/3}=0.001~\eV$ and $\omega_{ij}=0.5$ for the orthogonal $R$ matrix to evaluate BR($H^+\to \ell^+ N_{Ri}$). The results are shown in Table~\ref{Tab:BRHP} for both NH and IH, and will be employed for later signature simulations. From the table we are informed that a large hierarchy of individual branching ratios exists for both hierarchies with the largest being $\BR(H^+\to \tau^+ N_{R3})=0.343$ for NH and $\BR(H^+\to e^+ N_{R1})=0.358$ for IH.

\begin{table}[!htbp]
\begin{tabular}{|c|c|c|c|c|c|c|c|c|c|}
\hline
BR($H^+$) & $e^+ N_{R1}$ & $e^+ N_{R2}$ & $e^+ N_{R3}$ & $\mu^+ N_{R1}$ & $\mu^+ N_{R2}$ & $\mu^+ N_{R3}$ & $\tau^+ N_{R1}$ & $\tau^+ N_{R2}$ & $\tau^+ N_{R3}$
\\
\hline
NH& 0.023 & 0.004 & 0.050 & 0.163& 0.211 & 0.139 & 0.060 & 0.006 & 0.343
\\
\hline
IH& 0.358 & 0.015 & 0.109 & 0.005 & 0.192 & 0.027 & 0.009 & 0.197 & 0.088
\\
\hline
\end{tabular}
\caption{Branching ratios of the charged scalar $H^+$ into $\ell^+N_{Ri}$ for NH and IH at $m_{1/3}=0.001~\eV$ and $\omega_{ij}=0.5$.}
\label{Tab:BRHP}
\end{table}

\begin{table}[!htbp]
\begin{tabular}{|c|c|c|c|}
\hline
BR($H/A$) & ~$\nu N_{R1}$~ & ~$\nu N_{R2}$~ & ~$\nu N_{R3}$~
\\
\hline
NH & 0.246  & 0.221&  0.533
\\
\hline
IH & 0.372 & 0.404 &  0.224
\\
\hline
\end{tabular}
\caption{Branching ratios of the neutral scalars $H/A$ into $\nu N_{Ri}$ for NH and IH at $m_{1/3}=0.001~\eV$, $\omega_{ij}=0.5$ and upon summing over neutrinos $\nu$.}
\label{Tab:BRHA}
\end{table}

Concerning the neutral scalars $H$ and $A$, we sum over the light neutrinos when showing the scanning results, since they are invisible at colliders. In Fig.~\ref{BRHA}, BR($H/A\to \nu N_{Ri}$) is shown as a function of the lightest neutrino mass for NH and IH. No specific hierarchy in branching ratios is observed. A small difference between NH and IH is that the individual BR($H/A\to \nu N_{Ri}$) can exceed 0.5 in NH while it maximally reaches 0.5 in IH. The explicit values of $\BR(H/A\to \nu N_{Ri})$ are shown in Table~\ref{Tab:BRHA} using the same set of parameters as for $\BR(H^+\to \ell^+ N_{Ri})$. It is clear from the table that the largest individual branching ratio of $H/A$ is $\BR(H/A\to \nu N_{R3})=0.533$ for NH and $\BR(H/A\to \nu N_{R2})=0.404$ for IH. And all branching ratios of $H/A\to \nu N_{Ri}$ approach $1/3$ when the light neutrinos are nearly degenerate.

\subsection{Heavy Majorana Neutrinos}

\begin{figure}[!htbp]
\begin{center}
\includegraphics[width=0.45\linewidth]{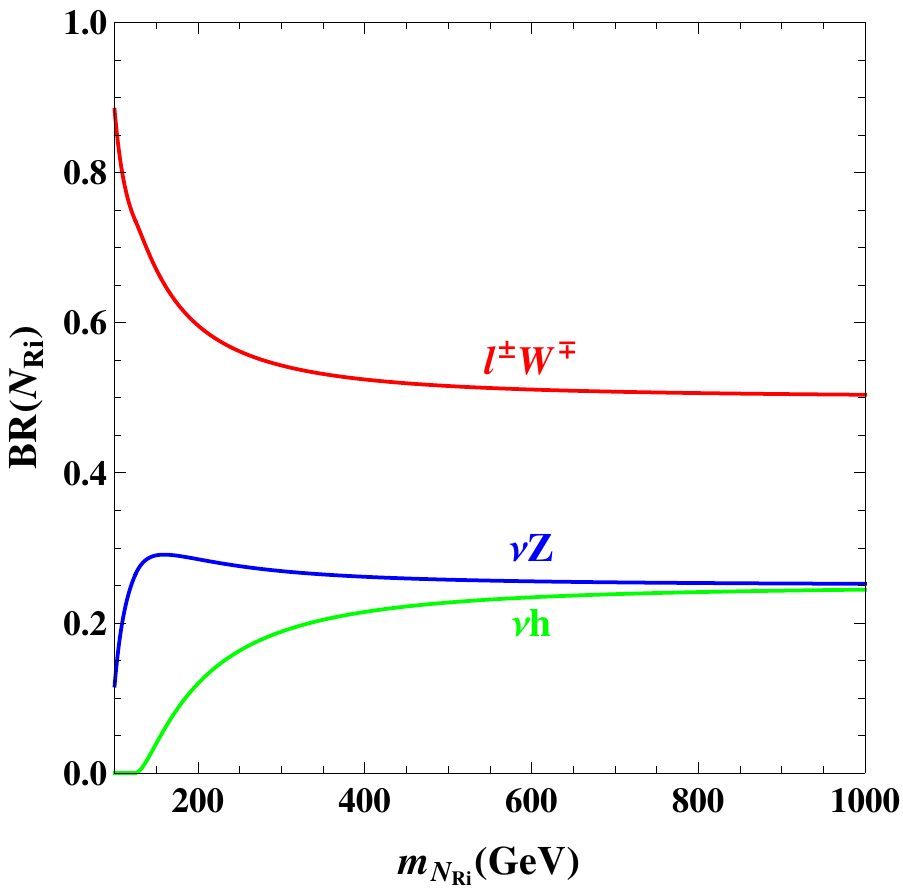}
\end{center}
\caption{Branching ratios of $N_{Ri}$ as a function of $m_{N_{Ri}}$ upon summing over the lepton flavors.}
\label{BRNR}
\end{figure}

When the heavy Majorana neutrinos are lighter than the neutrinophilic scalars, they decay via the small mixing with the light neutrinos. The partial decay widths are given by
\begin{eqnarray}
\Gamma(N_{Ri}\to\ell^\pm W^\mp)&=&\frac{|V_{\ell i}|^2}{8\pi v^2}m_{N_{Ri}}^3
\left(1-\Delta^W_{N_{Ri}}\right)^2\left(1+2\Delta^W_{N_{Ri}}\right),
\\
\Gamma(N_{Ri}\to \nu_\ell Z) &=& \frac{|V_{\ell i}|^2}{16\pi v^2}m_{N_{Ri}}^3
\left(1-\Delta^Z_{N_{Ri}}\right)^2\left(1+2\Delta^Z_{N_{Ri}}\right),
\\
\Gamma(N_{Ri}\to \nu_\ell h) &=& \frac{|V_{\ell i}|^2}{16\pi v^2}m_{N_{Ri}}^3
\left(1-\Delta^h_{N_{Ri}}\right)^2.
\end{eqnarray}
Due to the smallness of $V_{\ell N}$, $N_{Ri}$ are actually long-lived, which could lead to visible displaced vertices at LHC~\cite{Perez:2009mu}. In collider phenomenology study, the decay $N_{Ri}\to \ell^\pm W^\mp$ receives more attention, not only because it induces LNV signatures but also because it can be used to fully reconstruct the mass of $N_{Ri}$. In Fig.~\ref{BRNR}, we present the branching ratios of $N_{Ri}$ as a function of $m_{N_{Ri}}$ upon summing over the lepton flavors, which depend only on heavy neutrino masses $m_{N_{Ri}}$. As shown clearly, we have
\begin{equation}
\text{BR}(N_{Ri}\to \ell^\pm W^\mp):\text{BR}(N_{Ri}\to \nu Z):\text{BR}(N_{Ri}\to \nu h)
\approx 2:1:1,
\end{equation}
in the large $m_{N_{Ri}}$ limit, where the gauge bosons in the final state are mainly longitudinally polarized~\cite{Atre:2009rg}.

\begin{figure}
\begin{center}
\includegraphics[width=0.45\linewidth]{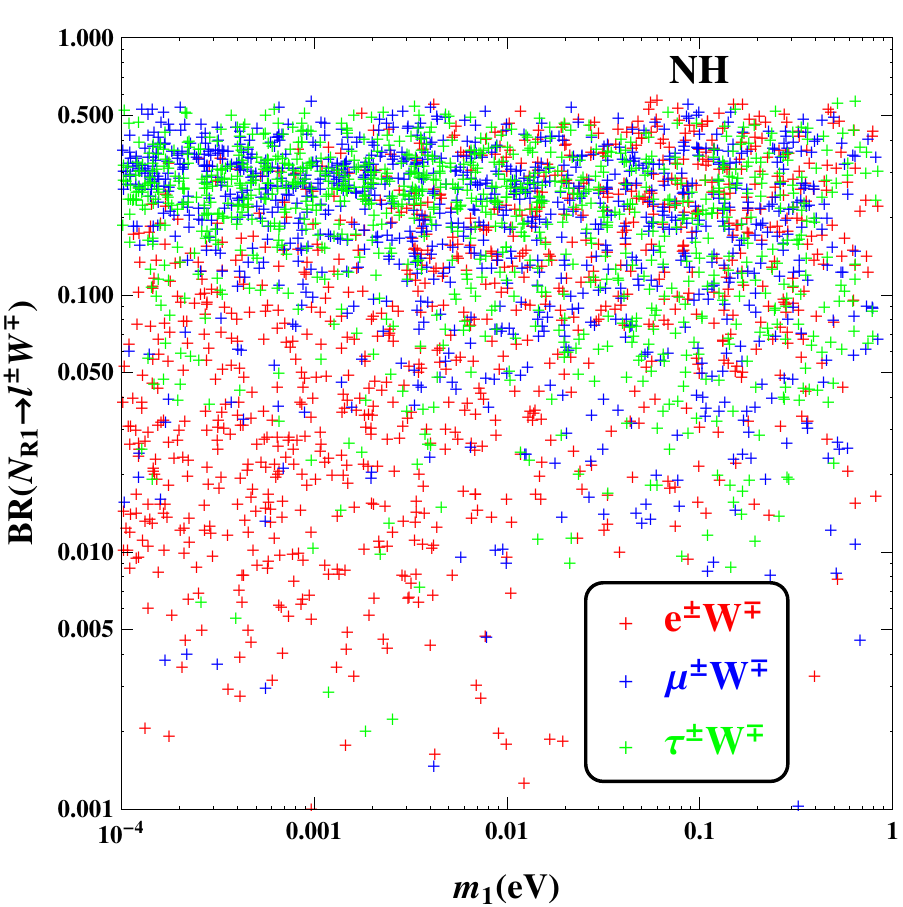}
\includegraphics[width=0.45\linewidth]{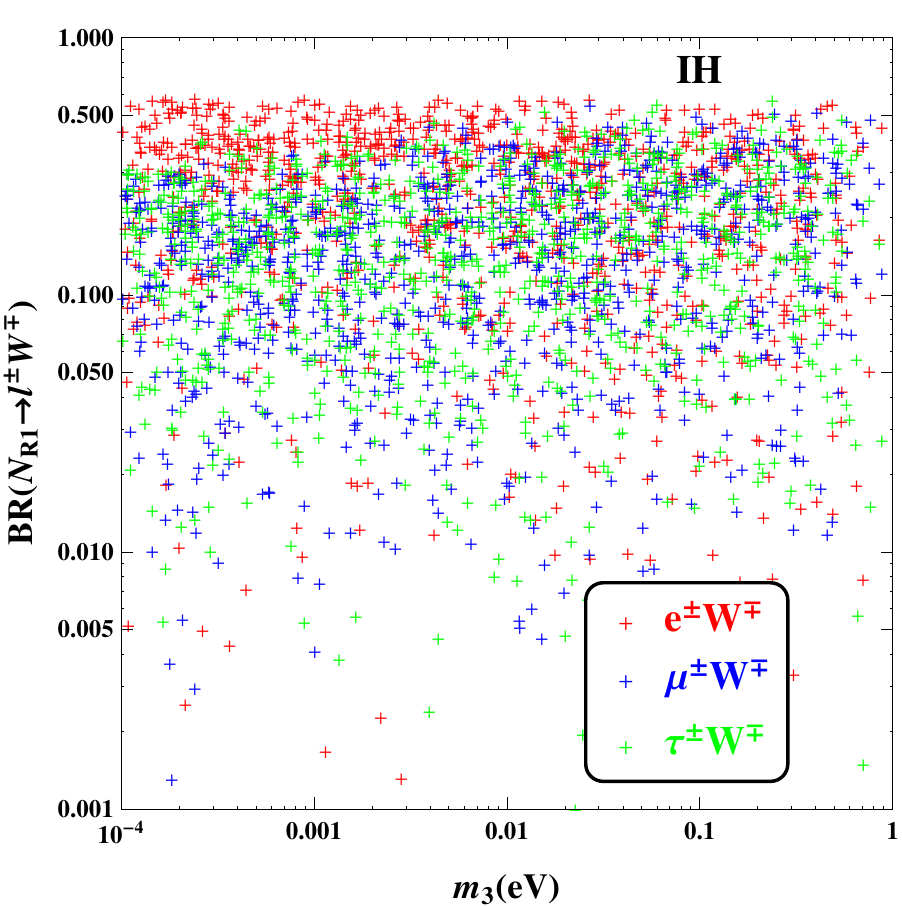}
\end{center}
\caption{Branching ratios of $N_{R1}$ into $\ell^\pm W^\mp$ as a function of the lightest neutrino mass for normal (left panel) and inverted (right) hierarchy at $m_{N_{R1}}=200~\GeV$.}
\label{BRNR1}
\end{figure}
\begin{figure}
\begin{center}
\includegraphics[width=0.45\linewidth]{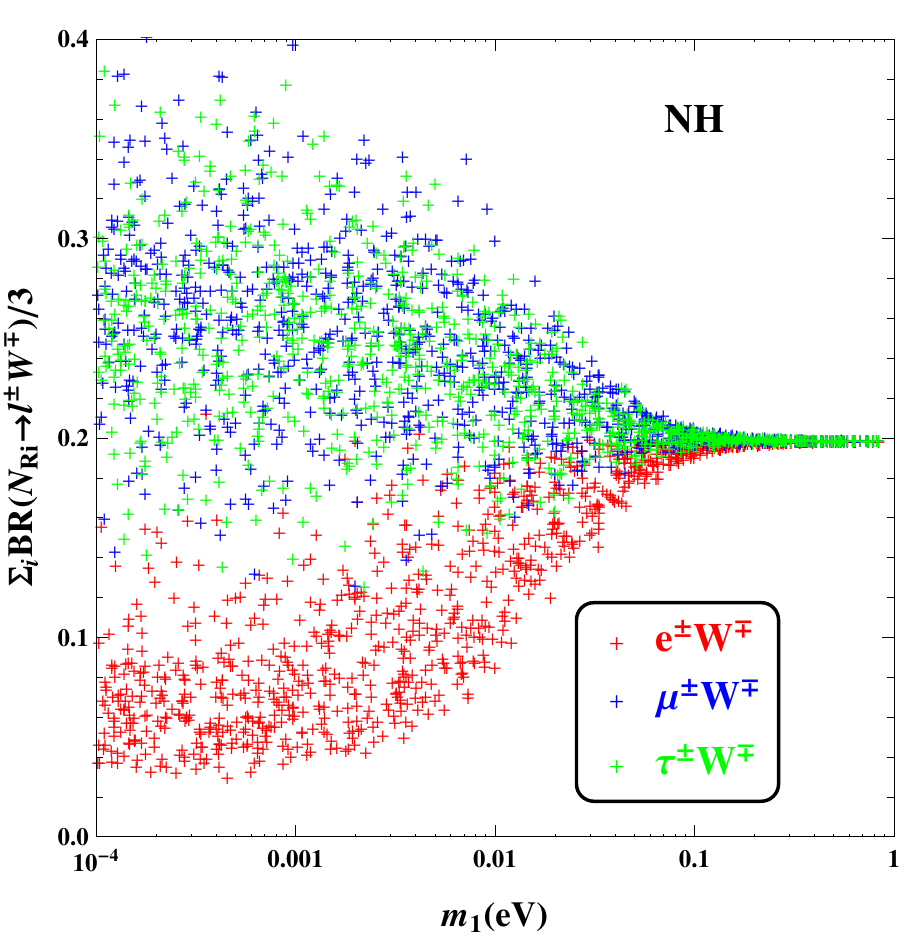}
\includegraphics[width=0.45\linewidth]{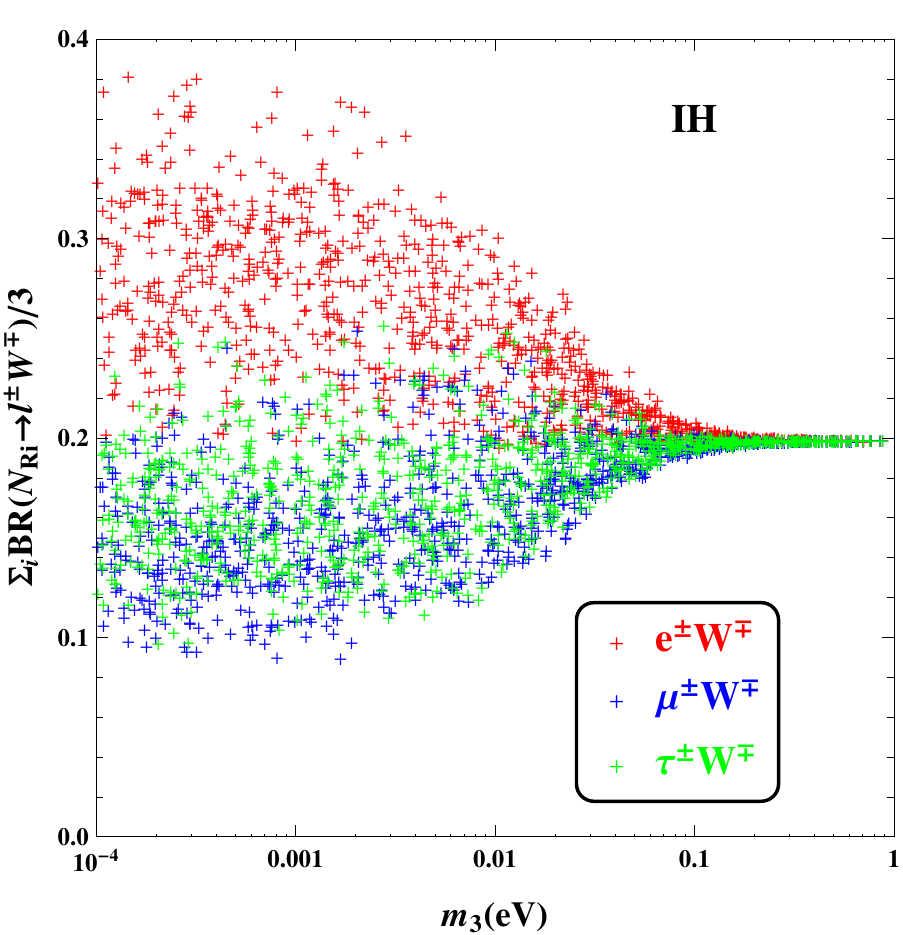}
\end{center}
\caption{The averaged branching ratios of $N_{Ri}$ as a function of the lightest neutrino mass for normal (left panel) and inverted hierarchy (right) at $m_{N_{Ri}}=200~\GeV$.}
\label{BRNRlW}
\end{figure}

For the decays into charged leptons there are similar flavor relations among $\BR(N_{Ri}\to\ell^\pm W^\mp)$ as in the decays of $H^+$. In Fig.~\ref{BRNR1}, the scanning results of $\BR(N_{R1}\to \ell^\pm W^\mp)$ are shown for $\ell=e,\mu,\tau$ separately as a function of the lightest neutrino mass at $m_{N_{R1}}=200~\GeV$, while the decays of $N_{R2}$ and $N_{R3}$ are similar. We observe the relations
\begin{eqnarray}
\label{HR}
\text{BR}(N_{R1}\to e^\pm W^\mp) &<& \text{BR}(N_{R1}\to \mu^\pm W^\mp)\approx\text{BR}(N_{R1}\to \tau^\pm W^\mp)~\text{for NH},\\
\text{BR}(N_{R1}\to e^\pm W^\mp) &>& \text{BR}(N_{R1}\to \mu^\pm W^\mp)\approx\text{BR}(N_{R1}\to \tau^\pm W^\mp)~\text{for IH},
\end{eqnarray}
when the lightest neutrino mass is less than $0.1~\eV$. These flavor relations can be better seen in Fig.~\ref{BRNRlW} which shows the results of the averaged branching ratios $\sum_i\BR(N_{Ri}\to\ell^\pm W^\mp)/3$. Hence the decays of heavy Majorana neutrinos $N_{Ri}$ into charged leptons could also be employed to distinguish between the neutrino mass hierarchies. Since $\BR(N_{Ri}\to \ell^\pm W^\mp)$ depends on $m_{N_{Ri}}$, we define the flavor ratio (FR)
\begin{equation}
\text{FR}(N_{Ri}\to \ell^\pm W^\mp)
= \frac{\text{BR}(N_{Ri}\to \ell^\pm W^\mp)}{\sum_\ell\text{BR}(N_{Ri}\to \ell^\pm W^\mp)}
= \frac{|V_{\ell i}|^2}{\sum_\ell |V_{\ell i}|^2},
\end{equation}
which is independent of $m_{N_{Ri}}$ but depends only on the neutrino oscillation parameters and the $R$ matrix. The distributions of $\textrm{FR}(N_{Ri}\to\ell^\pm W^\mp)$ and $\sum_i\textrm{FR}(N_{Ri}\to \ell^\pm W^\mp)/3$ are similar to Figs.~\ref{BRNR1} and \ref{BRNRlW} respectively, up to a normalization factor of $\sum_{\ell}\textrm{BR}(N_{Ri}\to \ell^\pm W^\mp)$. In Table~\ref{Tab:FRNR}, we show the values of $\textrm{FR}(N_{Ri}\to \ell^\pm W^\mp)$ for the same set of parameters as for $\BR(H^+\to \ell^+ N_{Ri})$. For NH, $N_{R1,2}\to \mu^\pm W^\mp$ and $N_{R3}\to \tau^\pm W^\mp$ are dominant while for IH $N_{R1}\to e^\pm W^\mp$, $N_{R2}\to \mu^\pm W^\mp/\tau^\pm W^\mp$, and $N_{R3}\to e^\pm W^\mp/\tau^\pm W^\mp$ take over. With the flavor ratios introduced above, we can easily acquire the branching ratios of $N_{Ri}\to \ell^\pm W^\mp$ for any values of $m_{N_{Ri}}$ by
\begin{equation}
\text{BR}(N_{Ri}\to \ell^\pm W^\mp)
=\text{FR}(N_{Ri}\to \ell^\pm W^\mp)\times\sum_\ell \text{BR}(N_{Ri}\to \ell^\pm W^\mp),
\end{equation}
where FR($N_{Ri}\to \ell^\pm W^\mp$) and $\sum_\ell \text{BR}(N_{Ri}\to \ell^\pm W^\mp)$ are given in Table~\ref{Tab:FRNR} and Fig.~\ref{BRNR}, respectively.

\begin{table}[!htbp]
\begin{tabular}{|c|c|c|c|}
\hline
FR($N_{Ri}$) & $e^\pm W^\mp$ & $\mu^\pm W^\mp$ & $\tau^\pm W^\mp$
\\
\hline
$N_{R1}$& 0.093 (0.962) & 0.663 (0.014)& 0.244 (0.024)
\\
\hline
$N_{R2}$& 0.019 (0.037)& 0.955 (0.476)& 0.026 (0.487)
\\
\hline
$N_{R3}$& 0.094 (0.486)& 0.261 (0.122) & 0.645 (0.392)
\\
\hline
\end{tabular}
\caption{Flavor ratios for heavy Majorana neutrinos $N_{Ri}$ decaying into $\ell^\pm W^\mp$ for NH (IH).}
\label{Tab:FRNR}
\end{table}

\begin{figure}
\begin{center}
\includegraphics[width=0.45\linewidth]{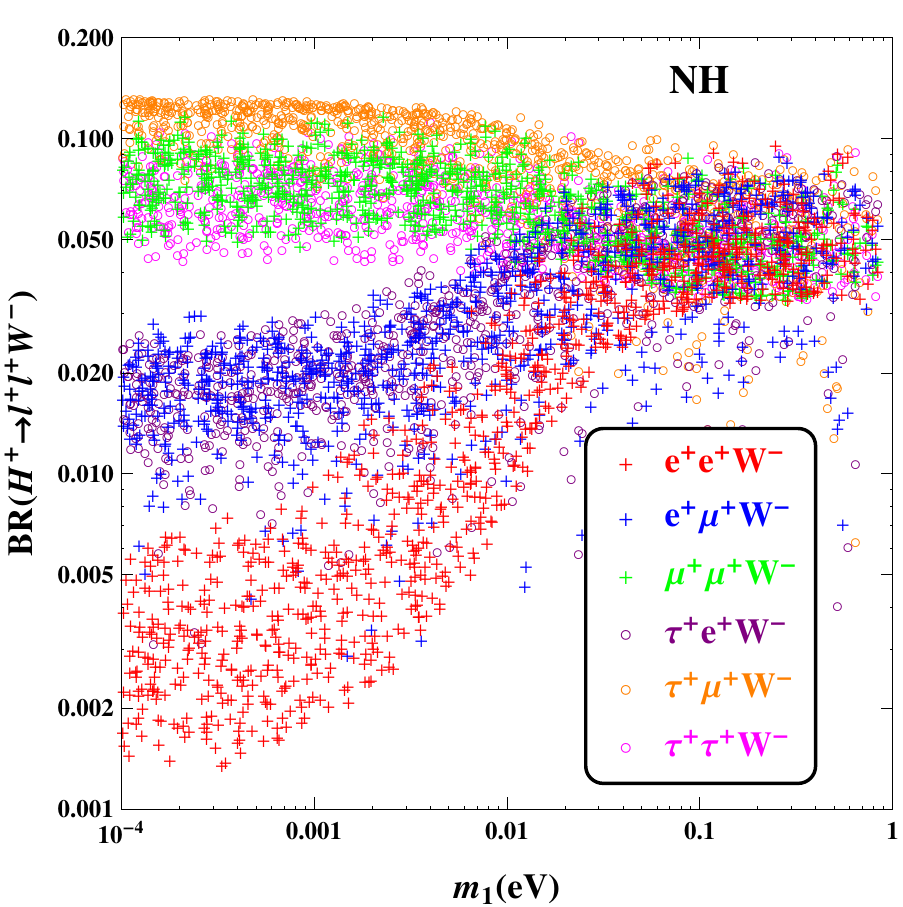}
\includegraphics[width=0.45\linewidth]{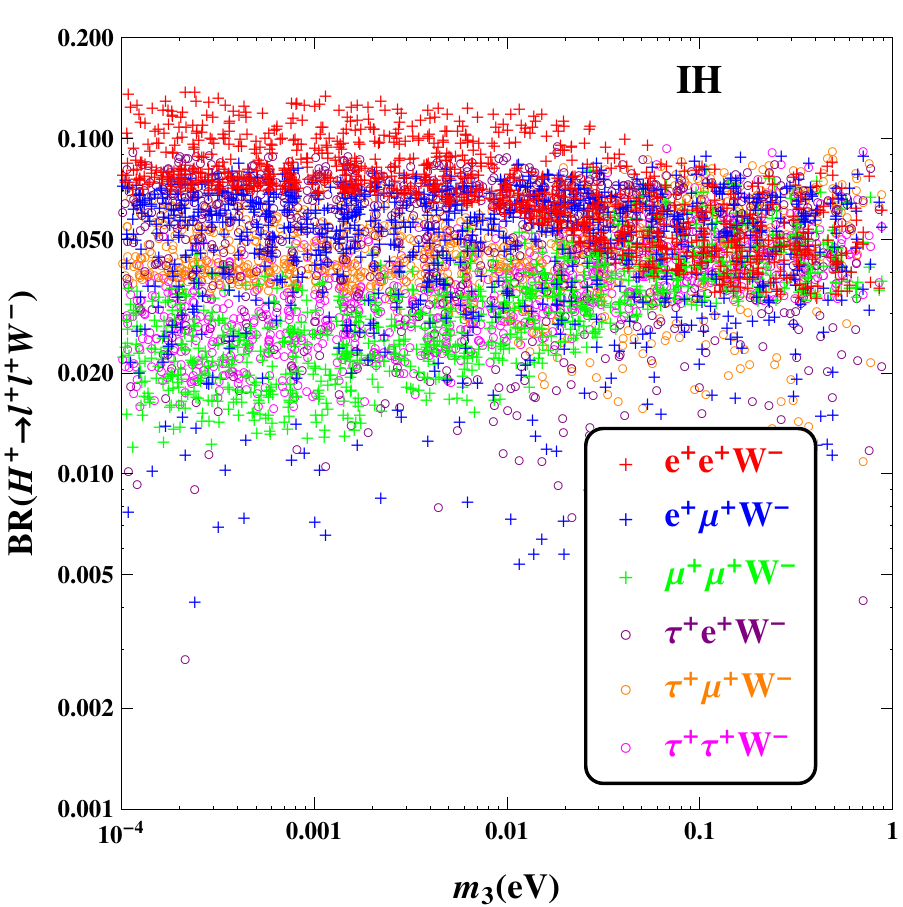}
\end{center}
\caption{Branching ratios of $H^+$ into $\ell^+\ell^+W^-$ as a function of the lightest neutrino mass for normal (left panel) and inverted (right) hierarchy at $(m_{H^+},m_{N_R})=(300,200)~\GeV$.}
\label{BRHPllW}
\end{figure}

Last but not least, we show in Fig.~\ref{BRHPllW} the scatter plots for $\BR(H^+\to \ell^+\ell^+ W^-)$ upon summing over the intermediate heavy Majorana neutrinos $N_{Ri}$. With $W^-$ further decaying hadronically, we have the LNV decay of the charged scalars $H^+\to \ell^+\ell^+jj$, which contributes to several LNV signatures at LHC. Since $\ell$ is identified as $e$ and $\mu$ at colliders, we have $\mu^+\mu^+$ dominance for NH and $e^+e^+$ dominance for IH when considering LNV signatures, which makes it possible to distinguish between the neutrino mass hierarchies.

\section{Lepton Number Violating Signatures}\label{SG}

\begin{figure}[!htbp]
\begin{center}
\includegraphics[width=0.45\linewidth]{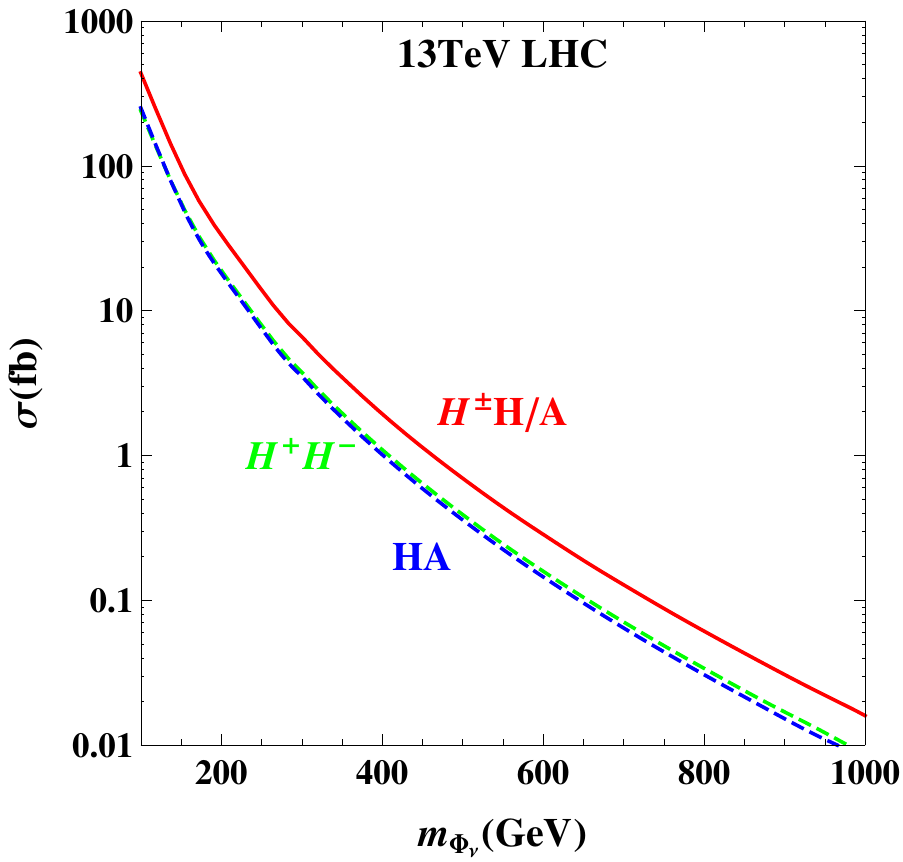}
\includegraphics[width=0.45\linewidth]{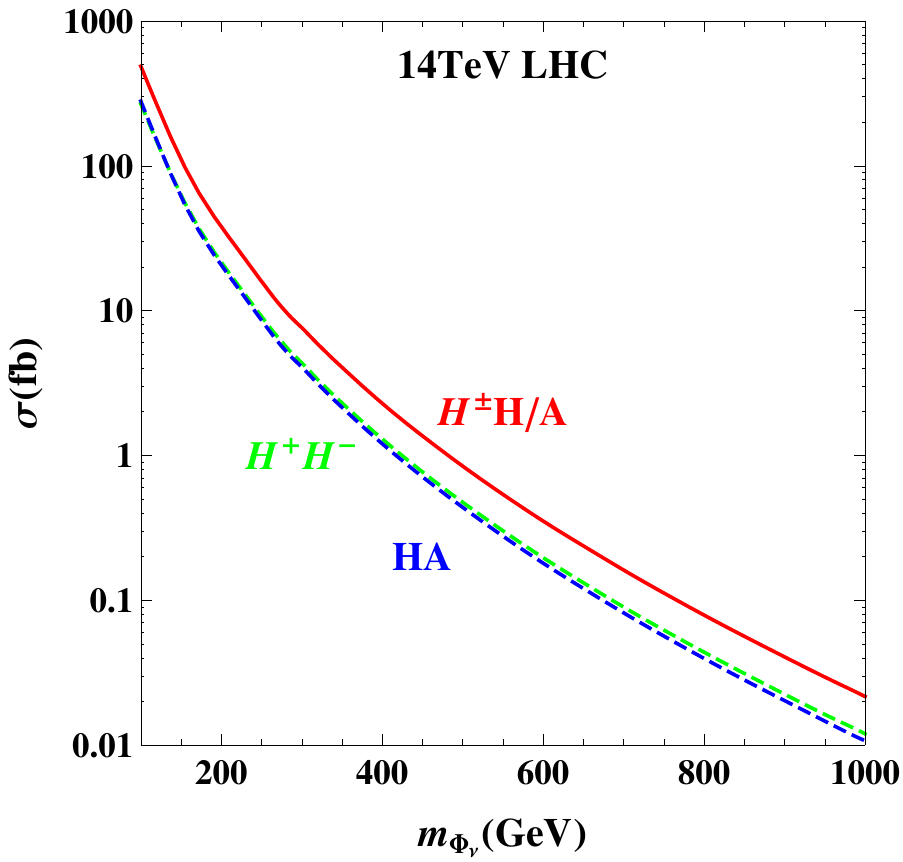}
\end{center}
\caption{Cross sections for the pair and associated production of the neutrophilic scalars at $13~(14)~\TeV$ LHC as a function of their mass $m_{H^+}\!=\!m_{H}\!=\!m_A\!=\!m_{\Phi_\nu}$.
\label{CS}}
\end{figure}

After the systematic study on the decay properties of the neutrinophilic scalars $H^\pm$, $H$, $A$ and the heavy Majorana neutrinos $N_{R}$ in Sec.~\ref{DP}, we can now investigate the LNV signatures at LHC. Our simulation procedure is as follows. We first implement the $\nu$2HDM into the package {\tt FeynRules}~\cite{Christensen:2008py} to generate the {\tt UFO}~\cite{Degrande:2011ua} model file. The parton level signal and corresponding background events are generated with {\tt MadGraph5\_aMC@NLO} \cite{Alwall:2011uj} using the {\tt NNPDF2.3} \cite{Ball:2012cx} LO parton distribution function set, and then pass through {\tt Pythia6} \cite{Sjostrand:2006za} to  include showering and hadronization. {\tt Delphes3}~\cite{Ovyn:2009tx} is then employed for a fast detector simulation and {\tt MadAnalysis5} \cite{Conte:2012fm} for analysis. Finally, the 95\% C.L. exclusion limits are acquired by employing {\tt CheckMATE}~\cite{Drees:2013wra}.

\begin{figure}
\begin{center}
\includegraphics[width=0.45\linewidth]{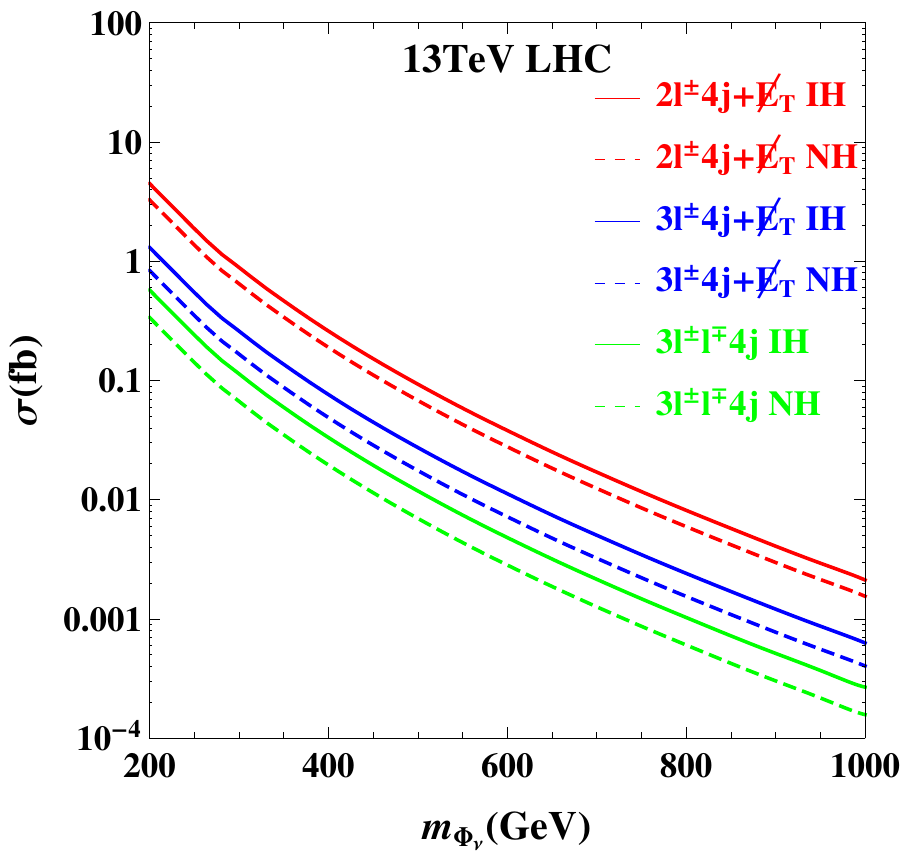}
\includegraphics[width=0.45\linewidth]{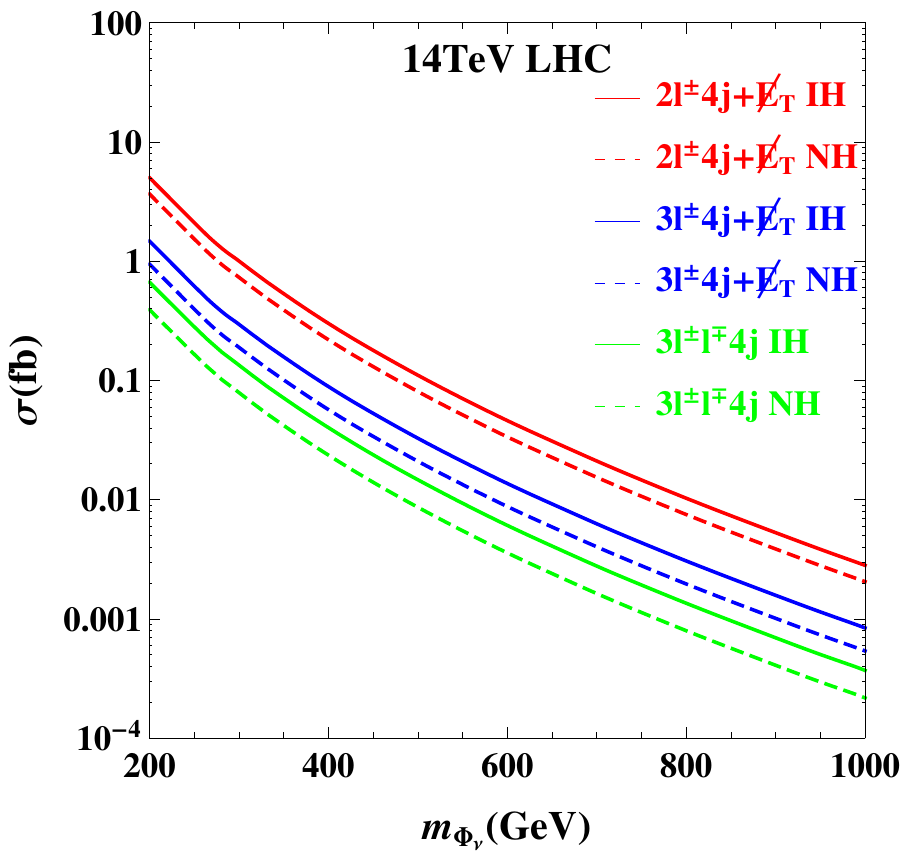}
\end{center}
\caption{Cross sections of LNV signatures at $13~(14)~\TeV$ LHC.
\label{CSLNV}}
\end{figure}

We first recall that the conventional well-studied LNV signature for heavy Majorana neutrinos, $pp\to W^\pm/H^\pm \to \ell^\pm N_R\to \ell^\pm\ell^\pm jj$, is also possible in $\nu$2HDM, but its amplitude is suppressed by $V_{\ell N}\sim10^{-7}$ and $v_\nu/v\sim 10^{-5}$ (for $v_\nu\sim\mathcal{O}(10~\MeV)$) respectively, making the signature practically unobservable at LHC. In contrast, due to the doublet nature of the neutrinophilic scalars, they can be pair and associated produced via the Drell-Yan processes
\begin{equation}
pp\to H^+H^-,~H^\pm H,~H^\pm A,~HA.
\end{equation}
Their cross sections at LHC are presented in Fig.~\ref{CS}, which range from $400$ to $0.01~\fb$ in the mass interval $100-1000~\GeV$ at 13 TeV, and become slightly enhanced at 14 TeV. There are many possible final states given by the decay channels of the scalars that we discussed in Sec.~\ref{DP} and the sequential decays of SM particles~\cite{Wang:2016vfj}. These channels lead to various signatures which are conventionally classified according to the multiplicities of charged leptons and jets. Among them the following three LNV signatures are most interesting and promising, and will be studied in Sec.~\ref{SSD}-\ref{subsec:four-lepton}:
\begin{itemize}
	\item $2\ell^\pm 4j + \cancel{E}_T$ from $H^\pm H$, $H^\pm A$ and $HA$ production,
	\item $3\ell^\pm 4j + \cancel{E}_T$ from $H^\pm H$, $H^\pm A$ production,
    \item $3\ell^\pm \ell^\mp 4j$ from $H^+H^-$ production,
\end{itemize}
where $\ell=e,\mu$ in our definition of a lepton for LHC signatures. In Fig.~\ref{CSLNV} we show the theoretical cross sections for the above LNV signatures at LHC. While the same sign dilepton (SSD) signature $2\ell^\pm 4j + \cancel{E}_T$ has the largest cross section, it is accompanied by relatively larger backgrounds. On the contrary, the four-lepton signature $3\ell^\pm \ell^\mp 4j$ is clean, but its cross section is also the smallest. In between, the same sign trilepton (SST) signature $3\ell^\pm 4j + \cancel{E}_T$ seems promising, since it is nearly background free. We also notice that the cross sections for all three signals are larger for neutrino masses in IH than in NH. To illustrate the testability of the LNV signatures, we choose the following benchmark points:
\begin{eqnarray}
\nonumber
\text{BP-A}:\ m_{N_R} &=& 200\GeV, m_{\Phi_\nu}=300\GeV,
\\
\nonumber
\text{BP-B}:\ m_{N_R} &=& 300\GeV, m_{\Phi_\nu}=400\GeV,
\\
\text{BP-C}:\ m_{N_R} &=& 400\GeV, m_{\Phi_\nu}=500\GeV,
\label{BP}
\end{eqnarray}
which are still allowed by current constraints. The signals and backgrounds will be simulated at $13~(14)~\TeV$ with an integrated luminosity of $100~(3000)~\fb^{-1}$, or LHC13@100 (LHC14@3000) for short, and the corresponding exclusion limits will be derived as well.

\subsection{Dilepton Signature}\label{SSD}

The signature comes from pair and associated production of the doublet scalar $\Phi_\nu$ and subsequent decays:
\begin{eqnarray}
\label{eq:ssd}
&&pp\rightarrow H^\pm H/A \rightarrow \ell^\pm N_R \nu  N_R \rightarrow \ell^\pm \ell^\pm W^\mp \nu \nu Z/h\rightarrow \ell^\pm \ell^\pm j j \nu \nu jj,
\\
&&pp\rightarrow H A \rightarrow \nu N_R \nu N_R\rightarrow \nu \ell^\pm W^\mp \nu \ell^\pm W^\mp \rightarrow \nu \ell^\pm jj \nu \ell^\pm jj,
\end{eqnarray}
where $\ell=e,\mu$ for collider simulations. The major sources of background are $t\bar{t}W, t\bar{t}Z$ and $W^\pm W^\pm W^\mp jj$. For the last one we use the {\tt MLM}~\cite{Hoche:2006ph} matching scheme with ${\tt xqcut}=25~(30)~\GeV$ for $13~(14)~\TeV$ LHC.

\begin{figure}
	\centering
	\includegraphics[scale=0.4]{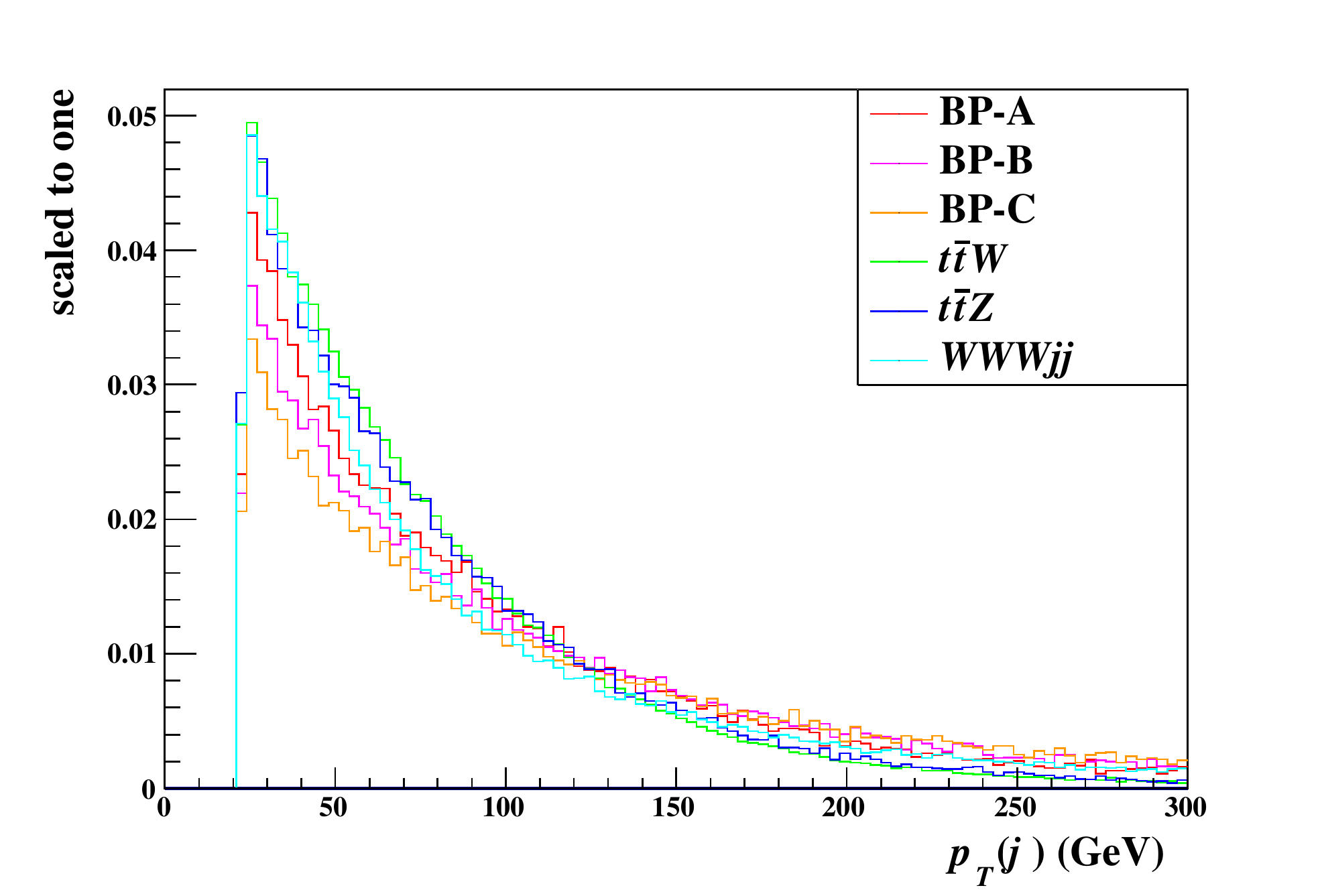}
	\includegraphics[scale=0.4]{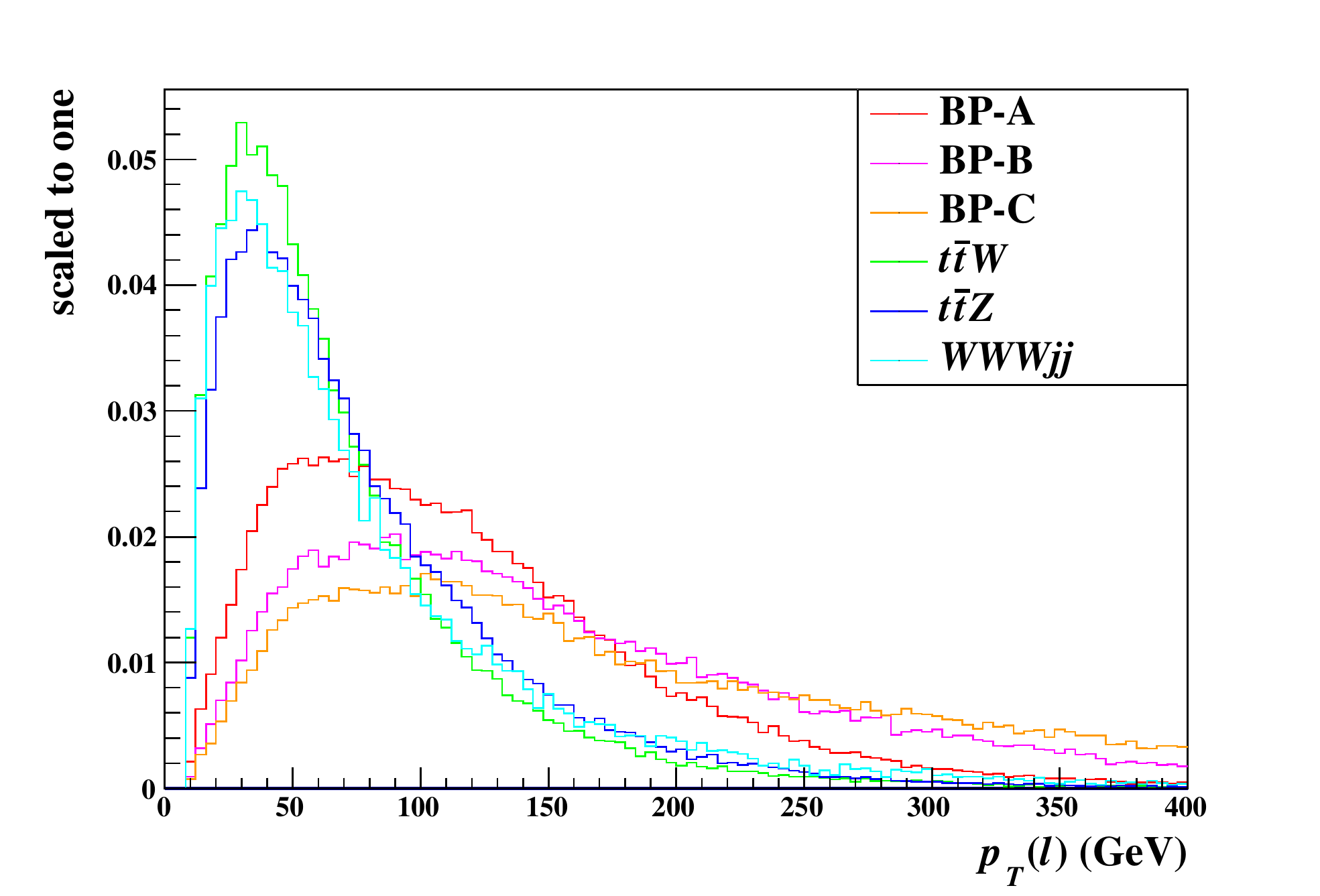}\\
	\includegraphics[scale=0.4]{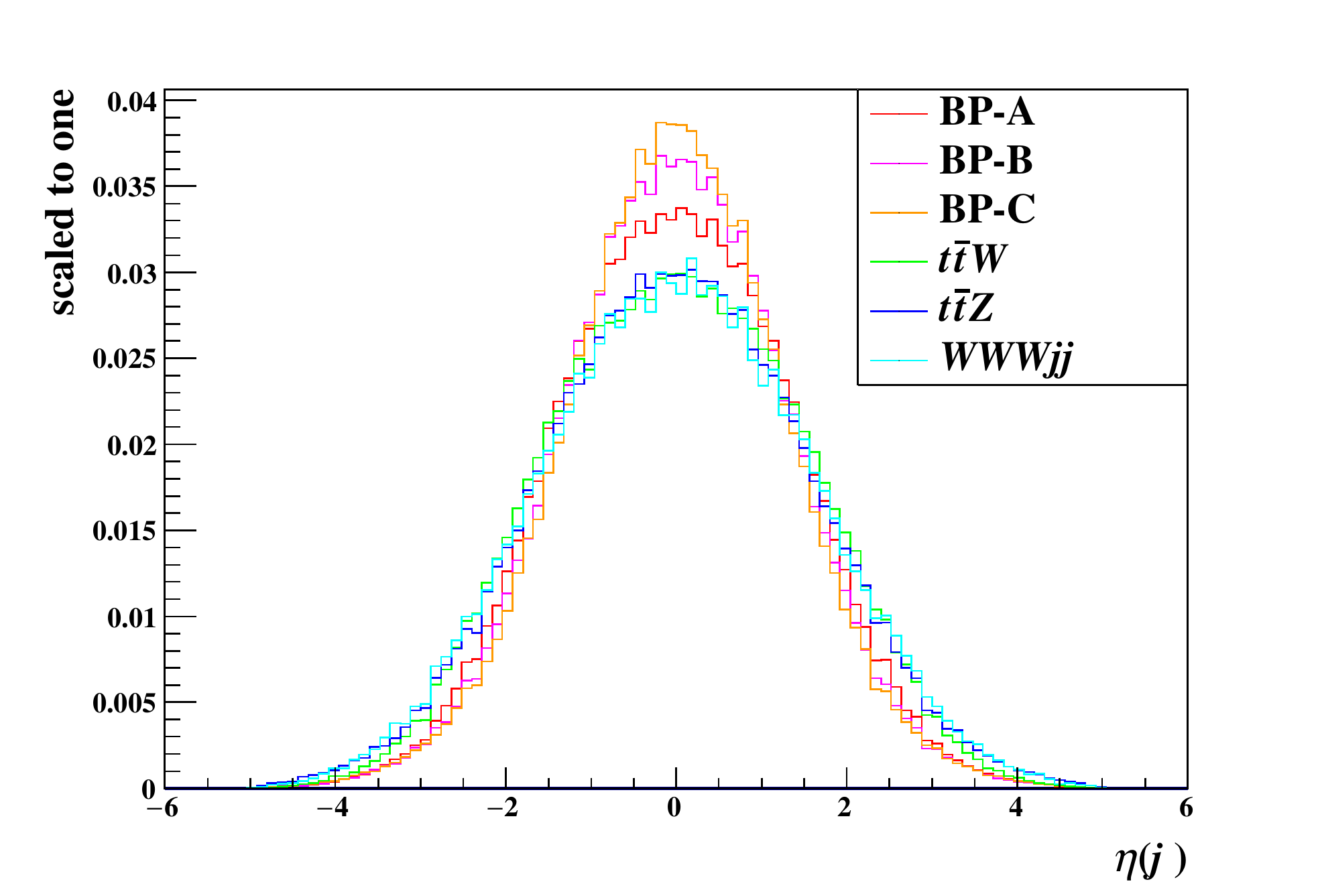}
	\includegraphics[scale=0.4]{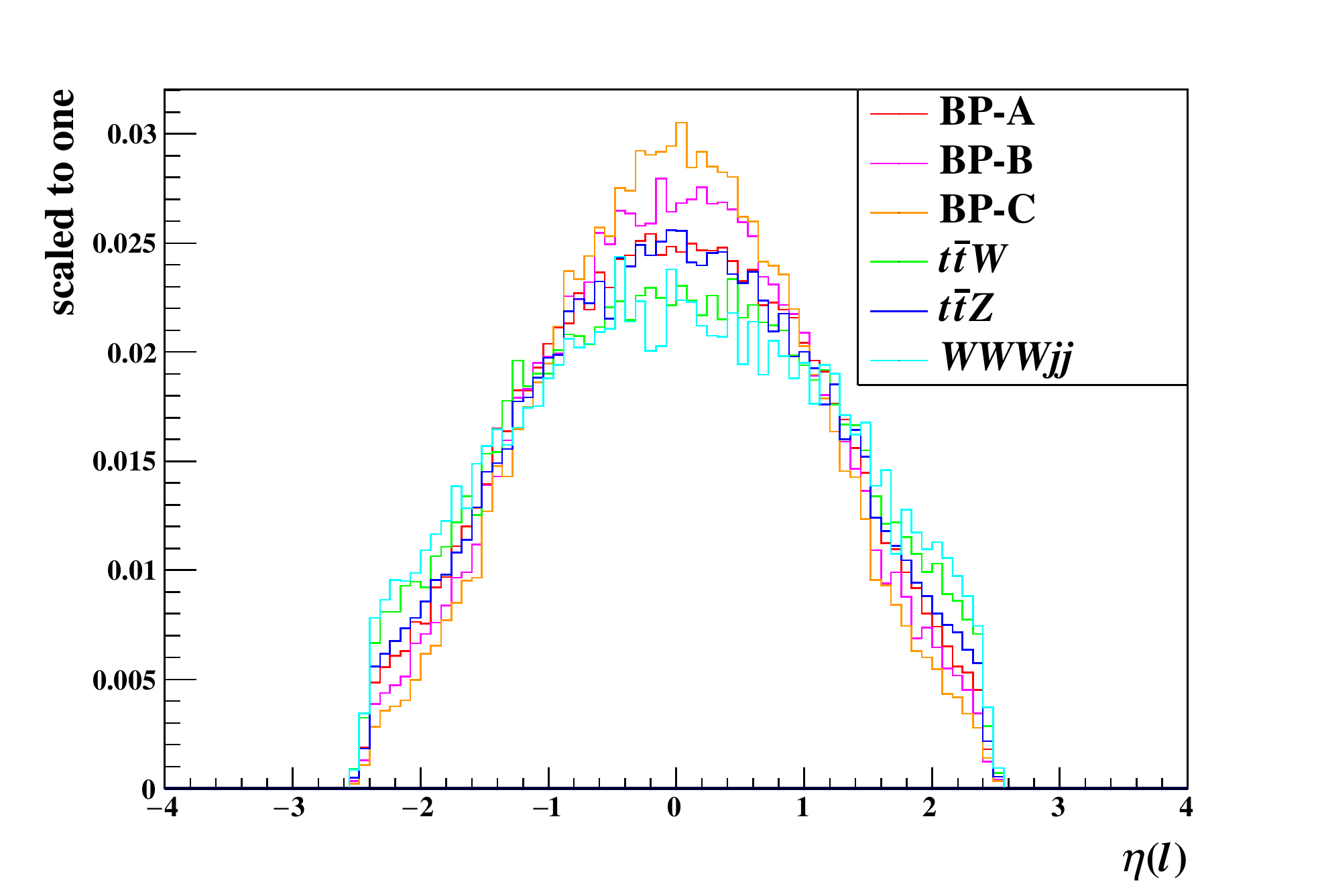}\\
    \includegraphics[scale=0.4]{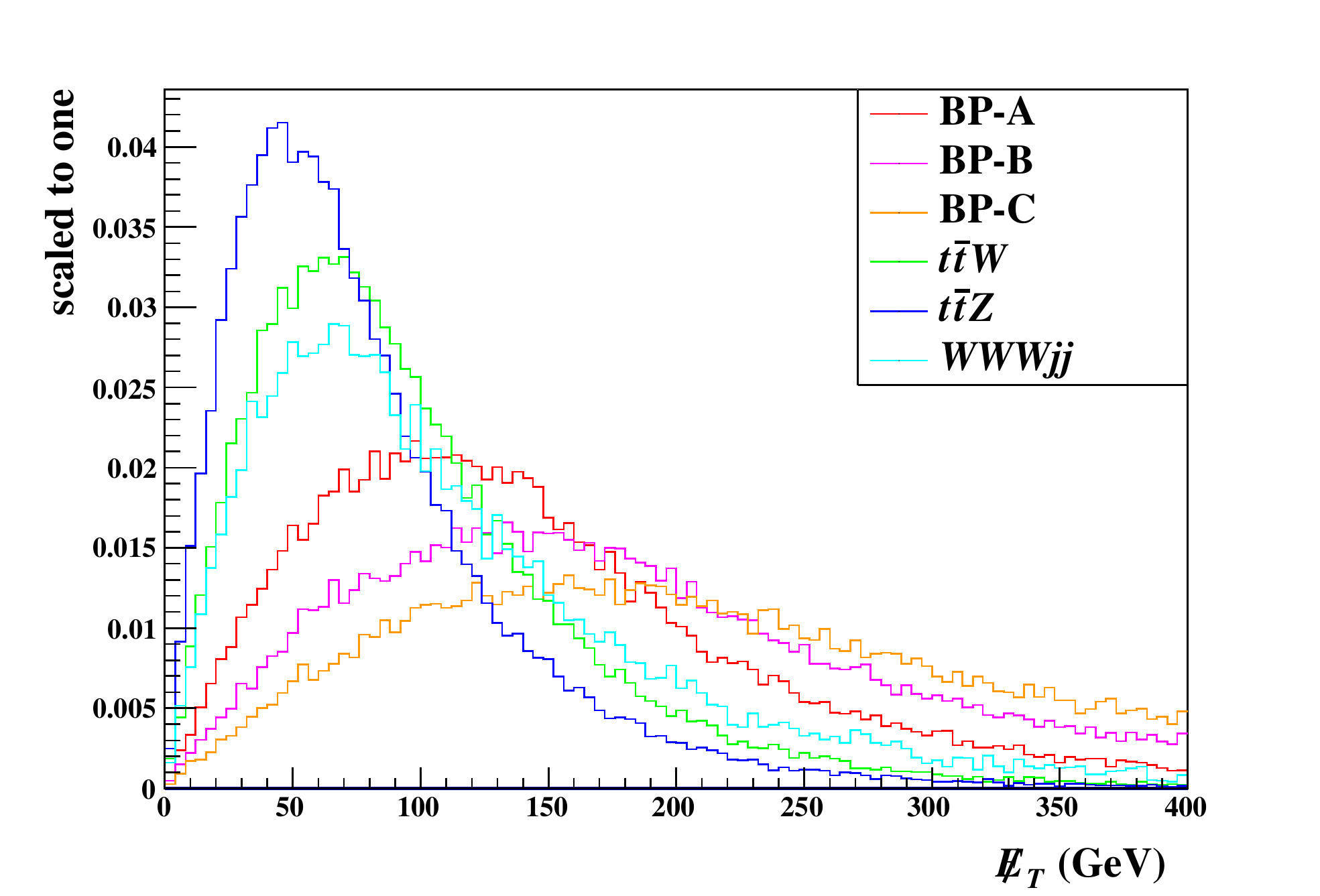}
	\includegraphics[scale=0.4]{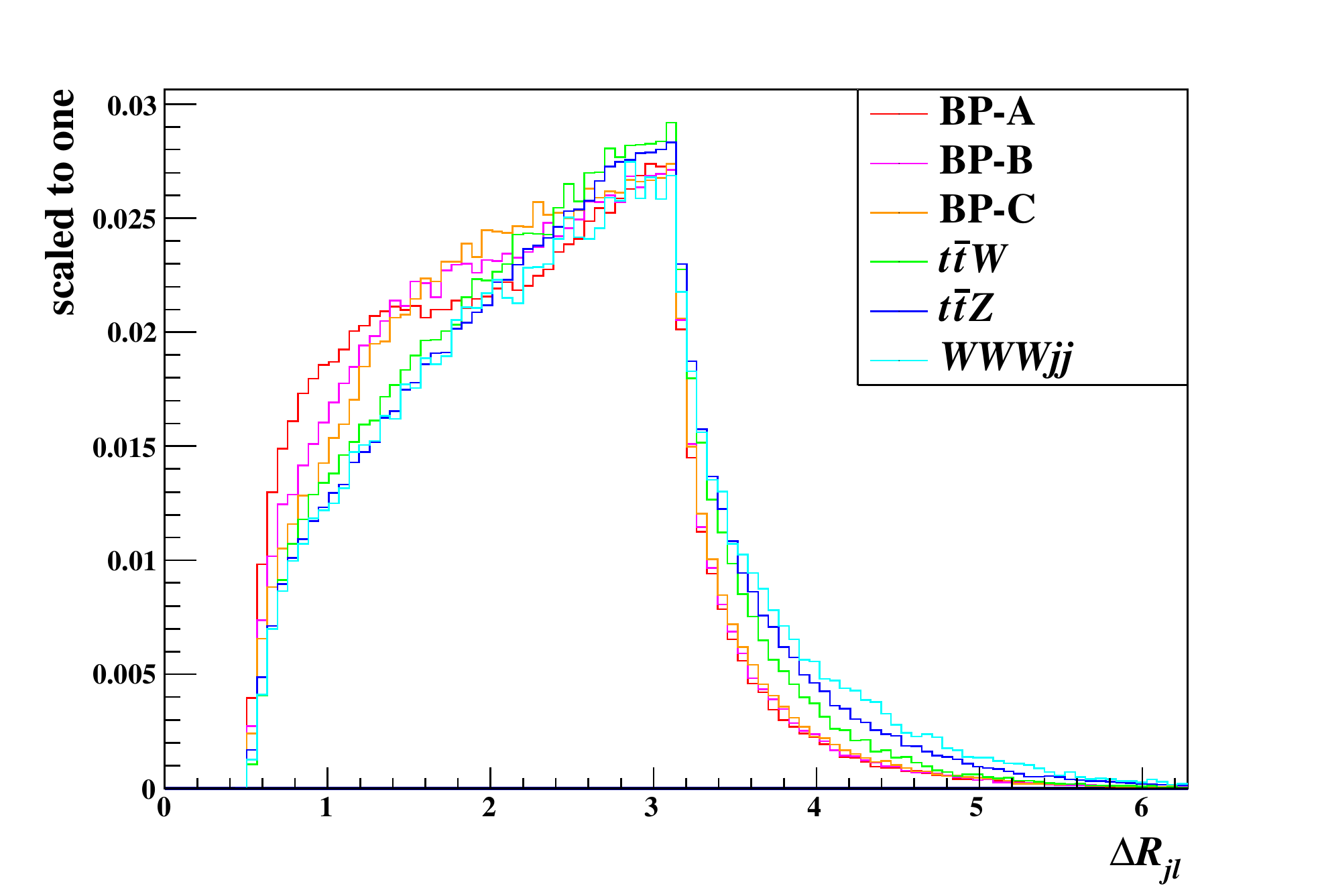}\\
	\includegraphics[scale=0.4]{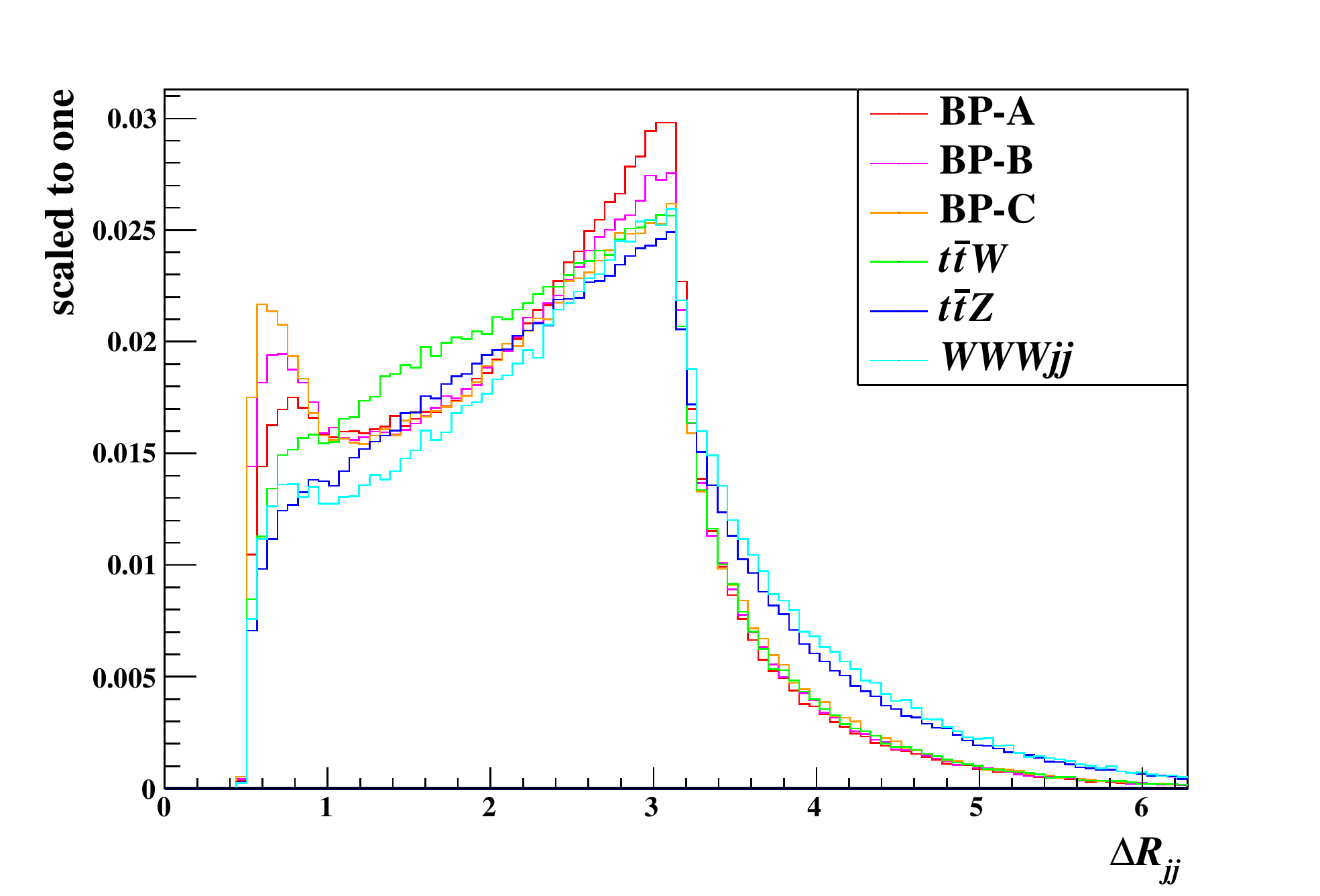}
	\includegraphics[scale=0.4]{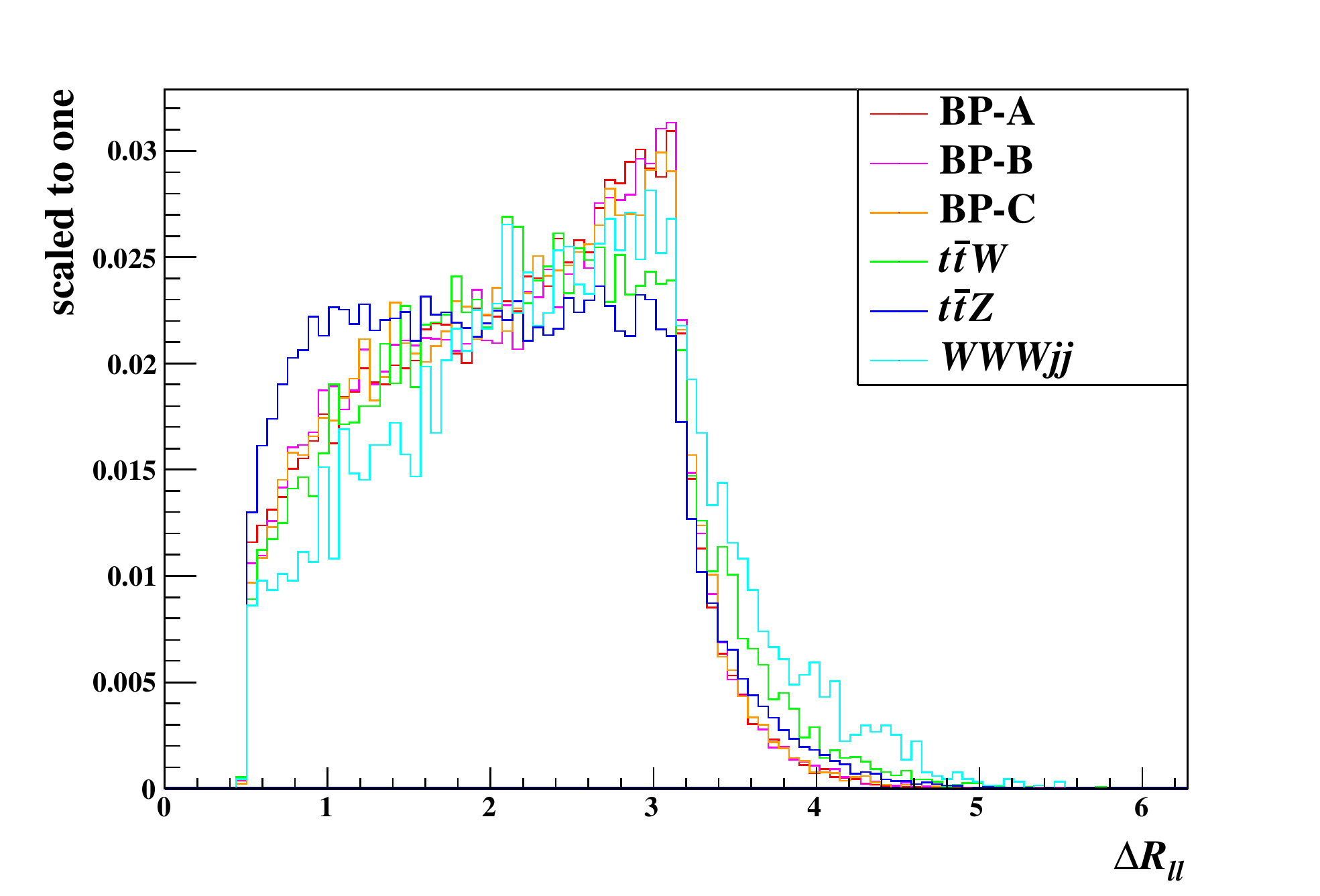}\vspace{-0.5em}
\caption{Distributions of transverse momenta $p_T(j),~p_T(\ell)$, pseudorapidities $\eta(j),~\eta(\ell)$, missing transverse energy $\cancel{E}_T$, and relative distances $\Delta R_{j\ell,jj,\ell\ell}$ for the SSD signature and corresponding backgrounds at 13 TeV LHC.}
	\label{fig:di_distributions_13}
\end{figure}

We show in Fig.~\ref{fig:di_distributions_13} the distributions of the transverse momentum $p_T(j),~p_T(\ell)$ and pseudorapidity $\eta(j),~\eta(\ell)$ for jets and leptons, the missing transverse energy $\cancel{E}_T$, and the relative distances $\Delta R_{j\ell,jj,\ell\ell}$ between leptons/jets for the SSD signature at 13 TeV LHC (and similarly at $14~\TeV$). Note that the normalized distributions for signatures from NH and IH are the same, despite the corresponding cross sections and lepton flavor structure are different. As the backgrounds are huge compared to the sigal, we only apply some basic cuts in order to preserve the signal to the maximum extent:
\begin{eqnarray}
p_T(\ell)>10\ \GeV,~\ p_T(j)>20\ \GeV,~\cancel{E}_T>10~\GeV,
\nonumber
\\
\label{eqn:basic cuts}
|\eta(\ell)|<2.5,\ |\eta(j)|<5,~\Delta R_{jj,\ell\ell,j\ell}>0.4.
\end{eqnarray}
In principle, we could tighten the cuts such as $p_T(\ell)>50~\GeV$, $\cancel{E}_T>100~\GeV$ to improve the signal to background ratio. But according to our simulation for the three benchmark points in Eq.~(\ref{BP}), such further cuts are actually not quite efficient to improve the significance, since the signal events are also suppressed heavily. Thus we only apply the following cuts to select the desired same-sign dilepton, four-jet events:
\begin{eqnarray}
\label{eqn:number cuts}
N(j)=4,~N(b)=0,~N(\ell^\pm)=2.
\end{eqnarray}
Here, the cut on the number of $b$-jet mainly aims to reduce the $t\bar{t}W$ and $t\bar{t}Z$ backgrounds. The identification of $b$-jets is performed with a tagging efficiency of $70\%$, a mis-tagging rate of $10\%$ for $c$-jets and $1\%$ for light-flavor jets, respectively \cite{Chatrchyan:2012jua}.

\begin{table}
	\begin{center}
		\begin{tabular}{|c|c|c|c|c|c|c|}
			\hline
			\multicolumn{2}{|c|}{Channels}    & Basic cuts in Eq:~(\ref{eqn:basic cuts})&
			$N(j)=4$   &    $N(b)=0$& $N(\ell^\pm)=2$&
			$S/\sqrt{S+B}$\\
			\hline
			\multirow{2}{*} {BP-A}
			 &NH  &  51 (1770)&  16 (569)& 8.3 (319)&  3.9 (164)& 0.65 (4.64) \\
			 & IH  &    63 (2425)&  20 (779)& 10 (436)&  4.9 (224)& 0.79 (6.20) \\
			\hline
			\multirow{2}{*}{BP-B}
			&NH    &    16 (566)&  5.0 (179)& 2.4 (90)&  1.3 (49)& 0.22 (1.44)\\
			&IH    &    20 (775)&  6.1 (245)& 3.0 (124)&  1.6 (67)& 0.27 (1.96)\\
			\hline
			\multirow{2}{*}{BP-C}
			&NH    &    5.8 (210)&  1.8 (65)& 0.87 (31)&  0.47 (17)& 0.08 (0.51)\\
			&IH    &    7.2 (288)&  2.2 (88)& 1.1 (43)&  0.58 (23)& 0.10 (0.70)\\
			\hline
			\multicolumn{2}{|c|}{$t\bar{t}W$}&   1020 (35023)& 325 (10917)& 54 (1810)& 18 (580)& --\\
			\multicolumn{2}{|c|}{$t\bar{t}Z$}&   1043 (37909)& 232 (8188) & 39 (1399)&  2.9 (172)&--\\
		     \multicolumn{2}{|c|}{$WWWjj$ }   &    155 (4623) &  43 (1213) & 32 (901) &  12 (334)&--\\
			\hline
		\end{tabular}
	\end{center}
\caption{Cut-flow for the SSD signature at three benchmark points in Eq.~(\ref{BP}) and dominant backgrounds at LHC13@100 (LHC14@3000) for both NH and IH.}\label{tab:dilepton_cuts}
\end{table}

\begin{figure}
	\centering
	\includegraphics[scale=0.4]{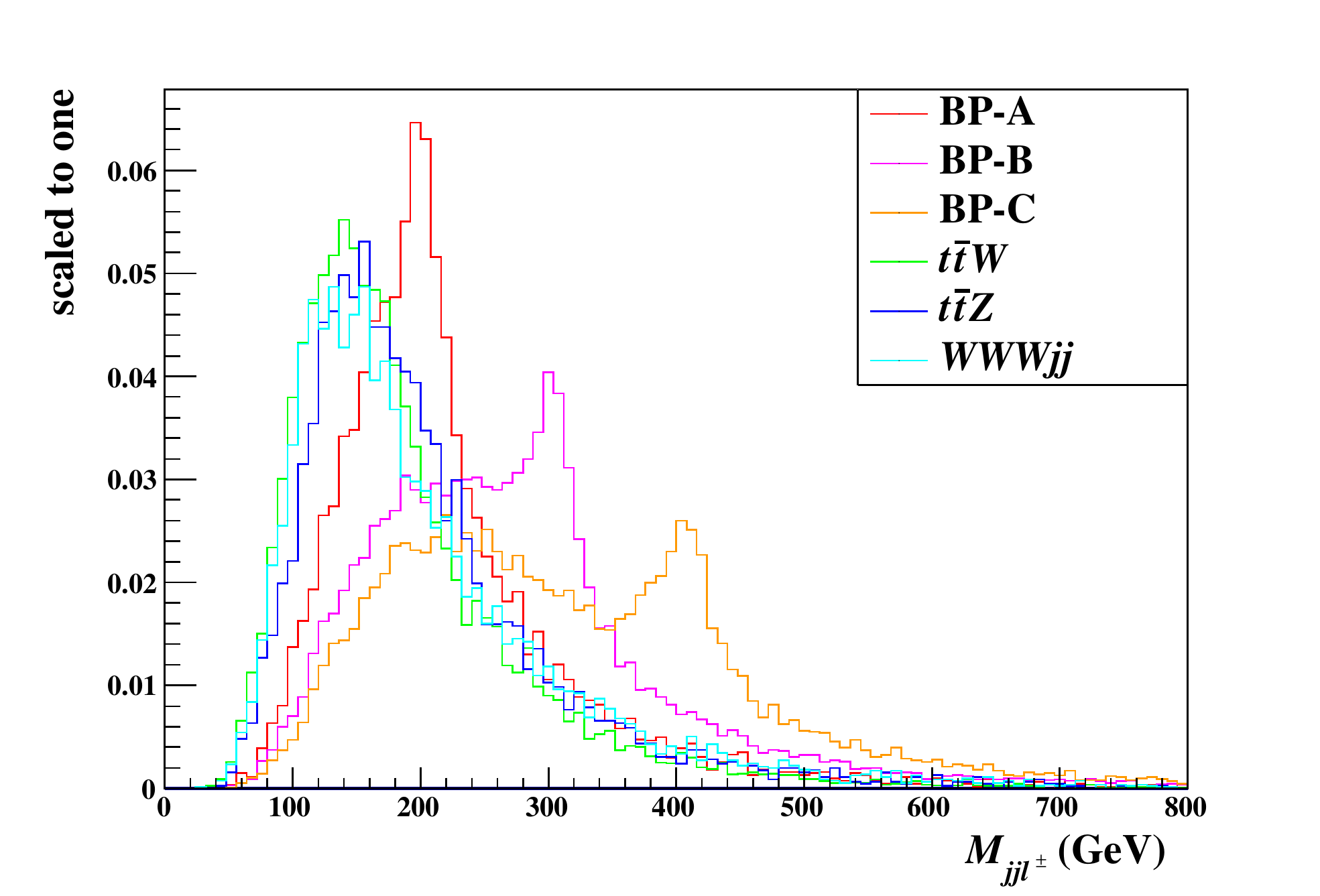}
	\includegraphics[scale=0.4]{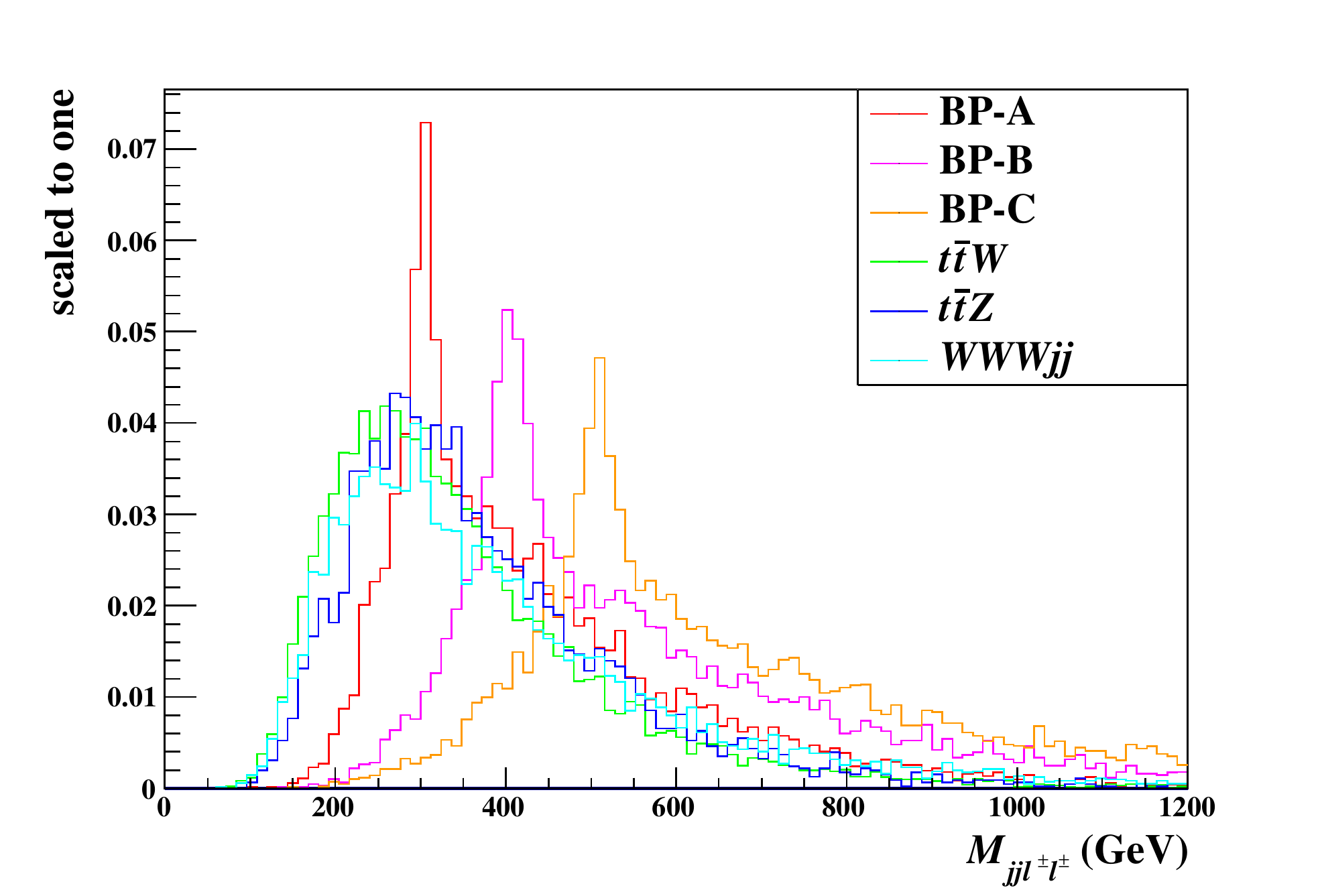}
\caption{Left (right): Reconstruction of $N_R$ ($H^\pm$) via invariant mass $M_{jj\ell^\pm}$ ($M_{jj\ell^\pm\ell^\pm}$) at $13~\TeV$ LHC for the SSD signature.}
	\label{fig:RC_13_di}
\end{figure}

In Table~\ref{tab:dilepton_cuts} we show the cut-flow for the SSD signature at the benchmark points and the dominant backgrounds, and list the statistical significance $S/\sqrt{S+B}$ in the last column, where $S$ ($B$) stands for the survival number of signal (background) events after applying all cuts. As shown in Fig.~\ref{fig:di_distributions_13}, the distributions of the background and signal events (especially for BP-A) are so similar for kinematic variables like $p_T^j$ and $\Delta R_{j\ell,\ell\ell}$ that it is hard to eliminate the background without sacrificing a big part of the signal events by the naive cuts in Eqs.~(\ref{eqn:basic cuts},\ref{eqn:number cuts}). As a result, the significance can barely reach $1\sigma$ at LHC13@100. However, at LHC14@3000, we may have a chance to probe the SSD signal for BP-A, even for which a more sophisticated and efficient cut strategy is highly desired to improve the testability of this SSD signature.

For the SSD signature the decay chains $N_R\to \ell^\pm jj$ and $H^\pm \to \ell^\pm\ell^\pm jj$ can be used to fully reconstruct the masses of heavy neutrinos $N_{Ri}$ and charged scalars $H^\pm$. In Fig.~\ref{fig:RC_13_di}, we depict the distributions for the reconstruction of $N_R$ and $H^\pm$ via the invariant mass $M_{jj\ell^\pm}$ and $M_{jj\ell^\pm\ell^\pm}$ respectively at $13~\TeV$ LHC after applying all cuts in Eqs.~(\ref{eqn:basic cuts},\ref{eqn:number cuts}). To render the resonance peaks more discernible, we have taken advantage of the expert mode of {\tt MadAnalysis5}~\cite{Conte:2012fm} so as to pick up close $jj\ell^\pm$ for $N_R$ and $jj\ell^\pm\ell^\pm$ for $H^\pm$ in the final states. We see that $N_R$ and $H^\pm$ are apparently reconstructible, although the production rate for BP-B and BP-C is actually too small to be detected even at LHC14@3000.

\begin{figure}
	\begin{center}
		\includegraphics[width=0.45\linewidth]{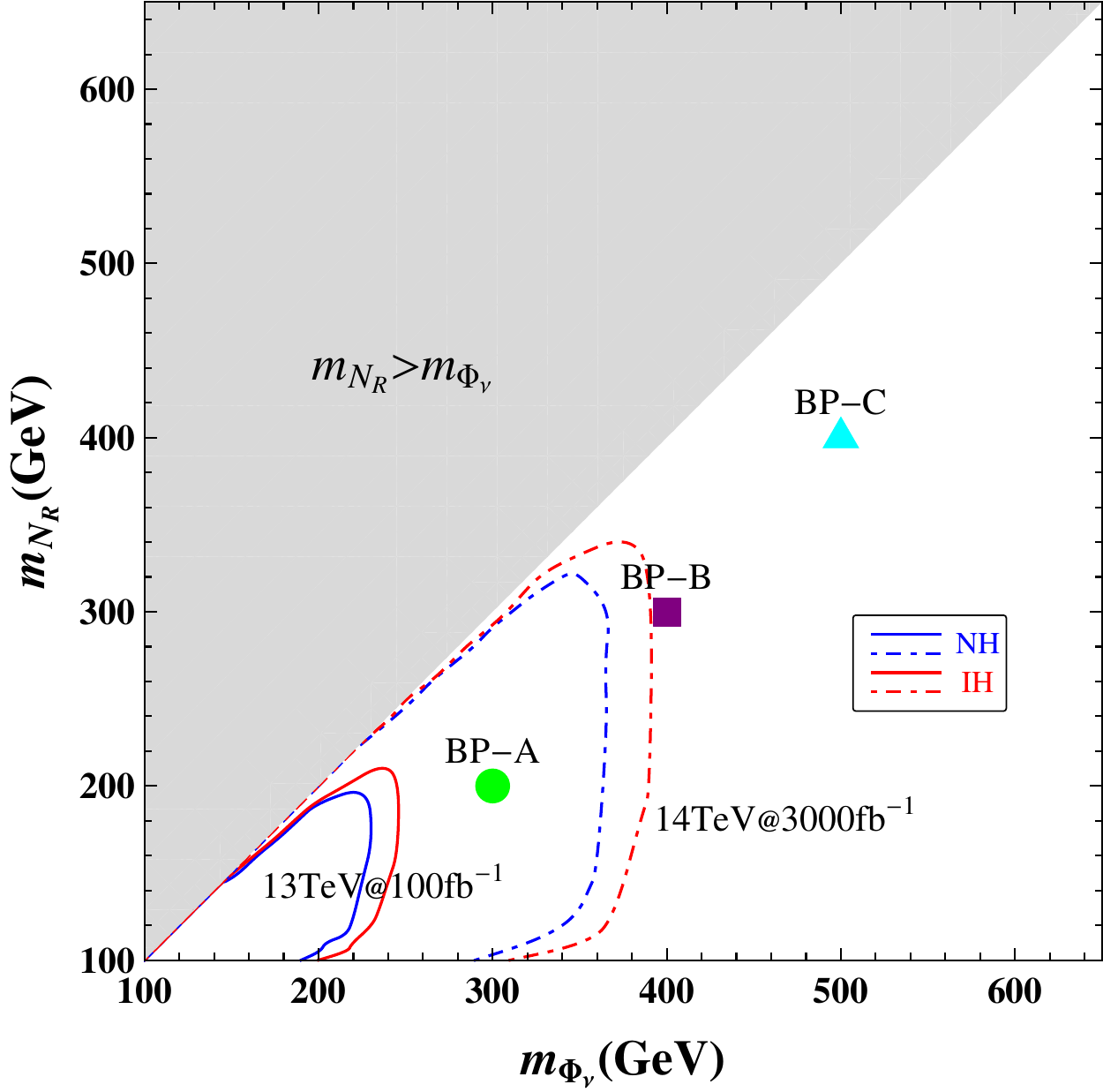}
	\end{center}
\caption{The expected 95\% C.L. exclusion limits in the $m_{N_R}-m_{\Phi_\nu}$ plane for the SSD signature at LHC13@100 and LHC14@3000.}
	\label{fig:dilepton_exclusion}
\end{figure}

Based on the above benchmark points study, it is worthwhile to figure out the exclusion limits of the SSD signature in the $m_{N_R}-m_{\Phi_\nu}$ plane. In Fig.~\ref{fig:dilepton_exclusion} we show the expected 95\% C.L. exclusion limits at LHC13@100 and LHC14@3000 by employing {\tt CheckMATE}~\cite{Drees:2013wra} with the cuts in Eqs.~(\ref{eqn:basic cuts},\ref{eqn:number cuts}). We see that while LHC13@100 rules out a parameter region with $m_{N_R}\lesssim 200~\GeV$ and $m_{\Phi_\nu}\lesssim 250~\GeV$, LHC14@3000 can exclude a larger region up to $m_{N_R}\lesssim 350~\GeV$ and $m_{\Phi_\nu}\lesssim 400~\GeV$. Nevertheless, a compressed spectrum with $m_{\Phi_\nu}\approx m_{N_R}$ is still allowed for both NH and IH, because due to the limited phase space for the $\Phi_\nu\to N_R$ decay the momentum of the leptons or $\cancel{E}_T$ is too small to pass the basic cuts in Eq.~(\ref{eqn:basic cuts}). We also note that the limits are more stringent for IH than NH. This originates simply from the fact that the cross section for the SSD signature in the IH case is about $1.37$ times as large as the one in the NH case as shown in Fig.~\ref{CSLNV}. As also shown clearly in Fig.~\ref{fig:dilepton_exclusion}, although all three benchmark points are beyond the reach of LHC13@100, they are either within the reach (BP-A), on the edge (BP-B) or out (BP-C) of the exclusion capability of LHC14@3000.

\subsection{Trilepton Signature}

\begin{figure}
	\centering
	\includegraphics[scale=0.4]{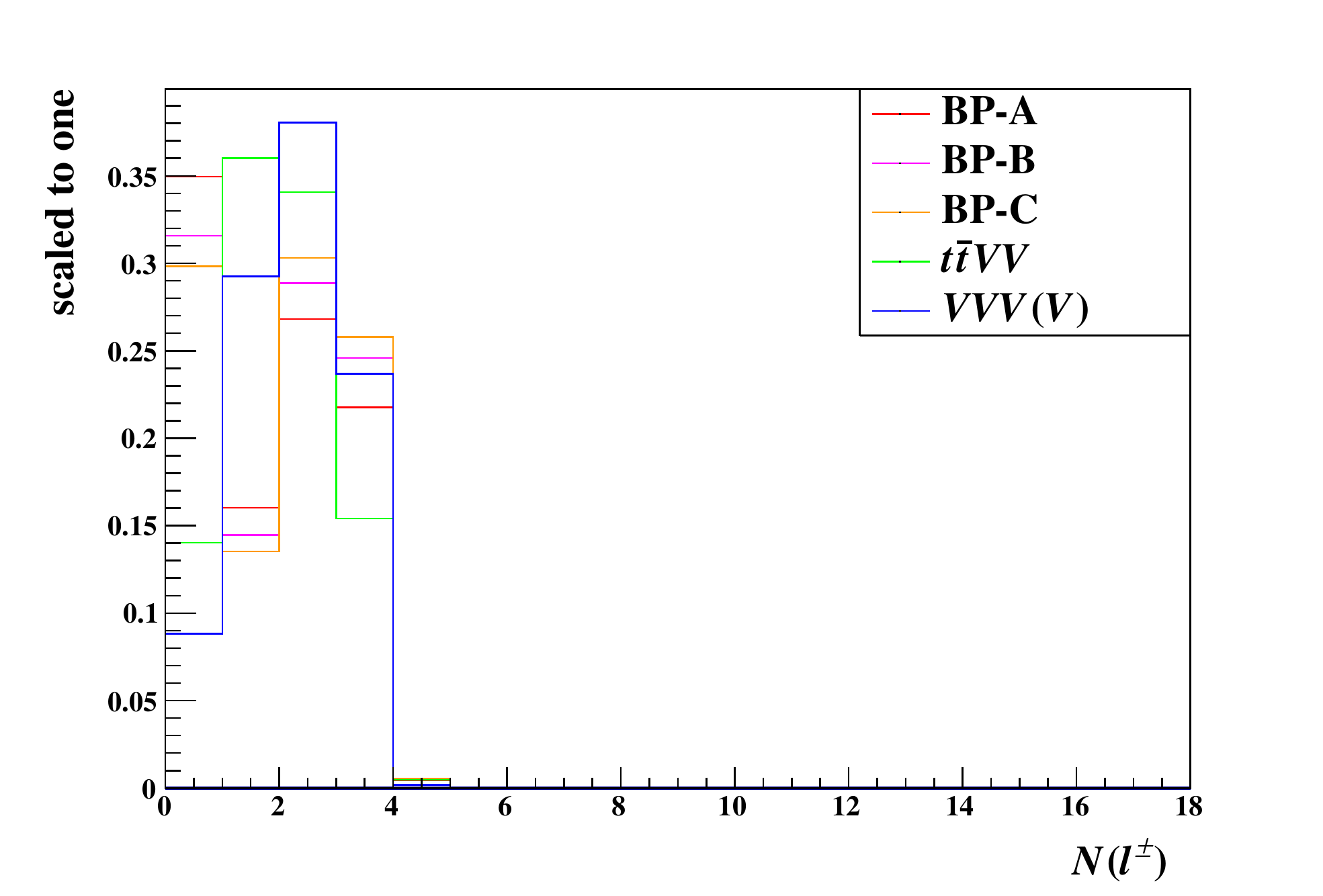}
	\includegraphics[scale=0.4]{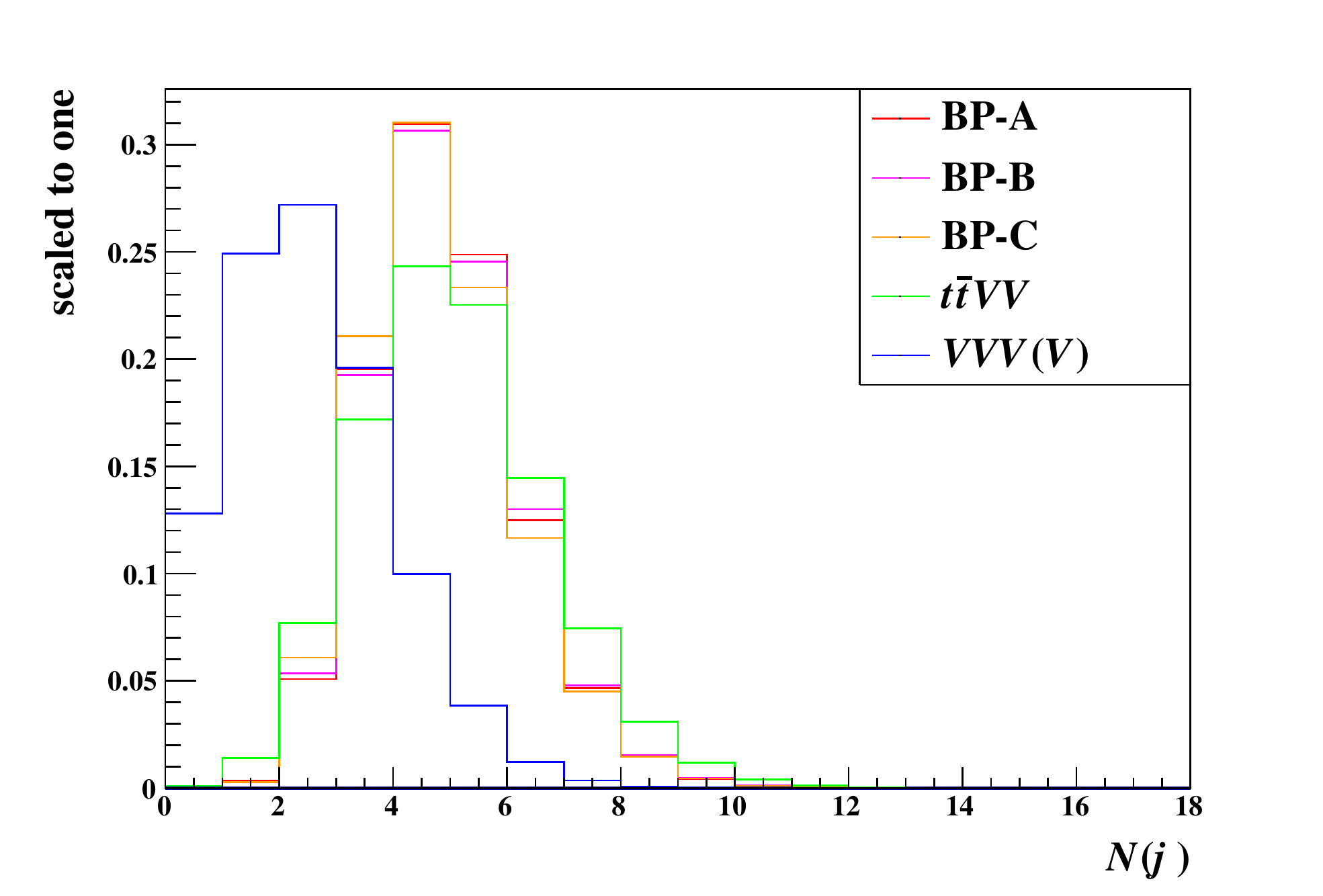}\\
	\includegraphics[scale=0.4]{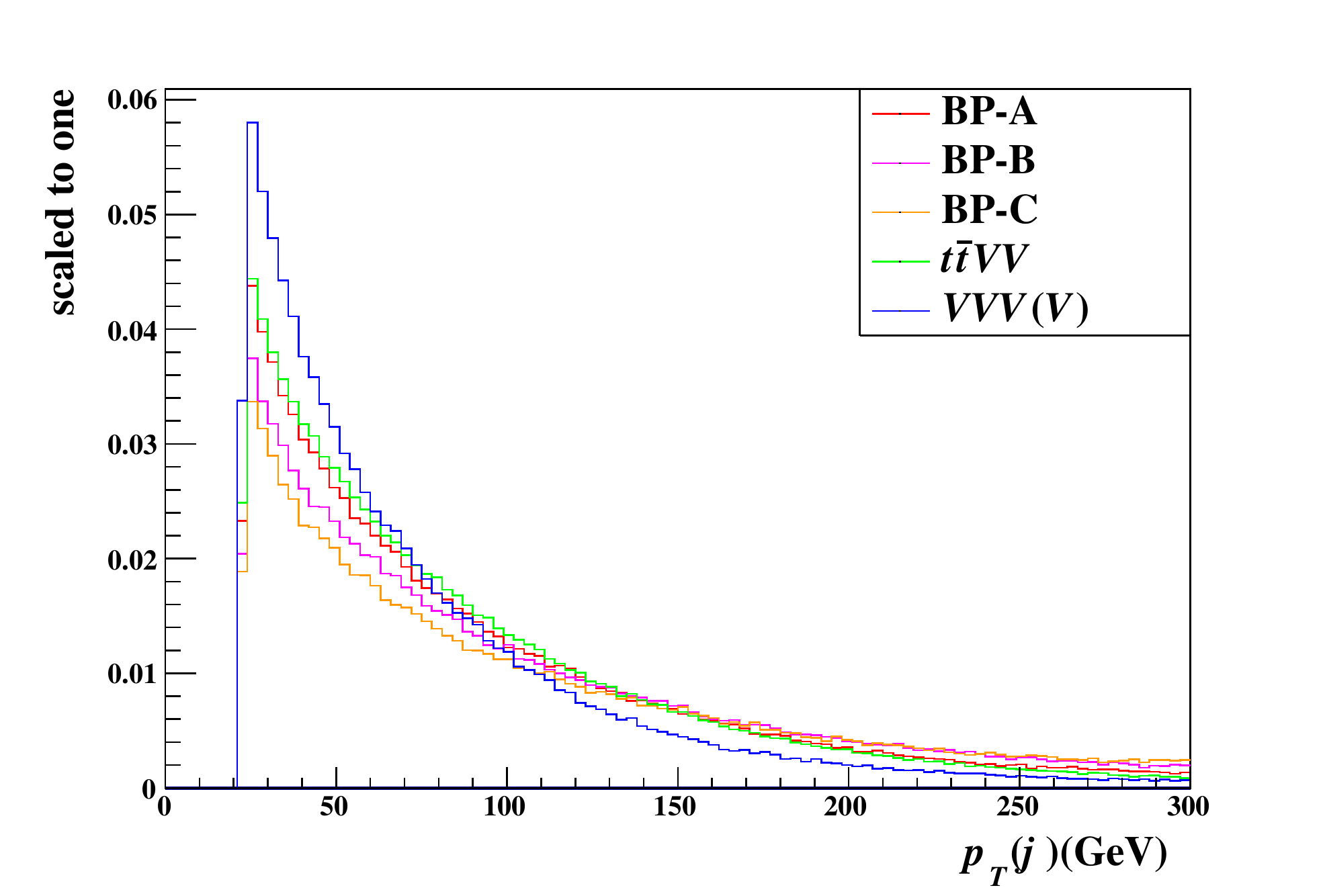}
	\includegraphics[scale=0.4]{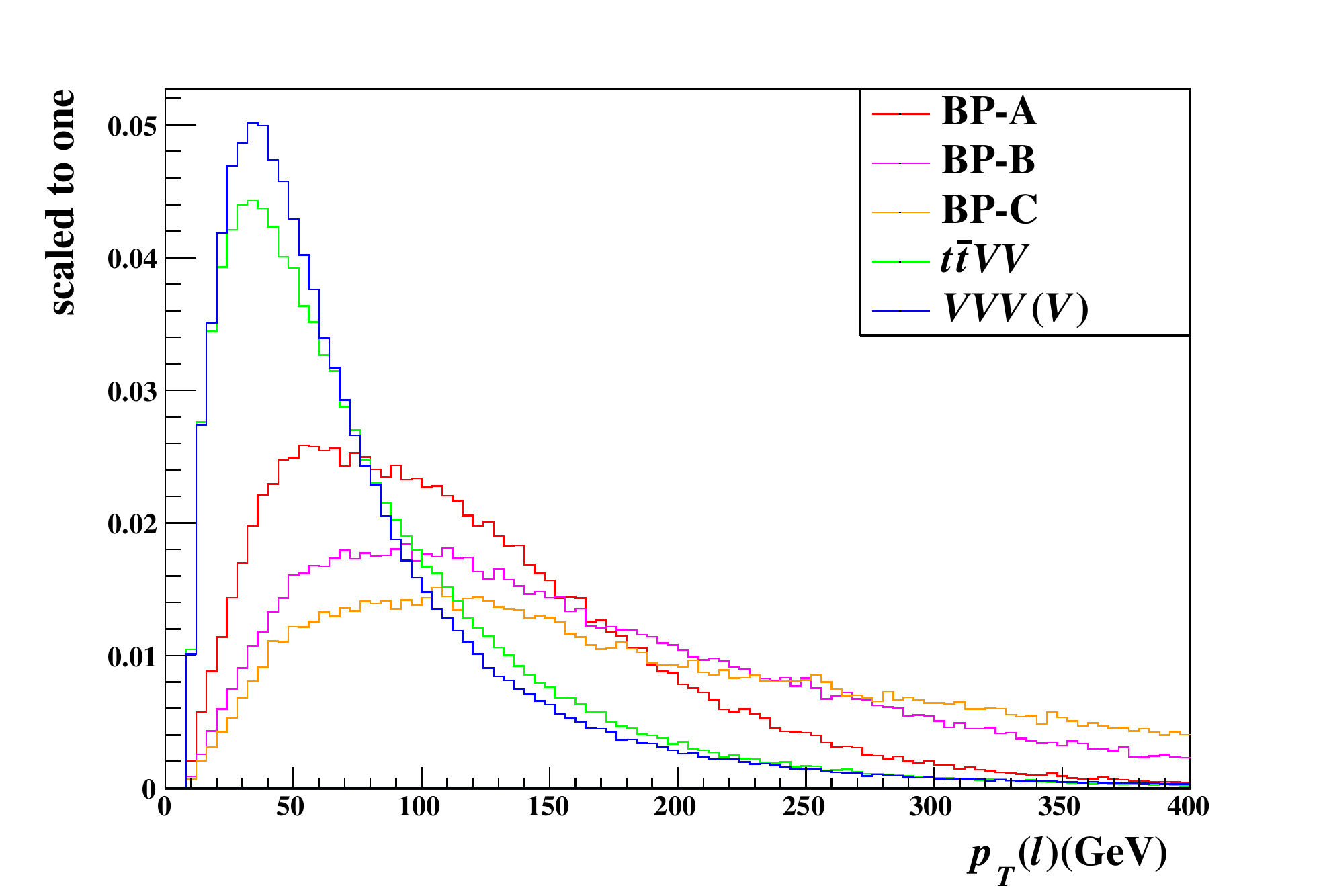}\\
    \includegraphics[scale=0.4]{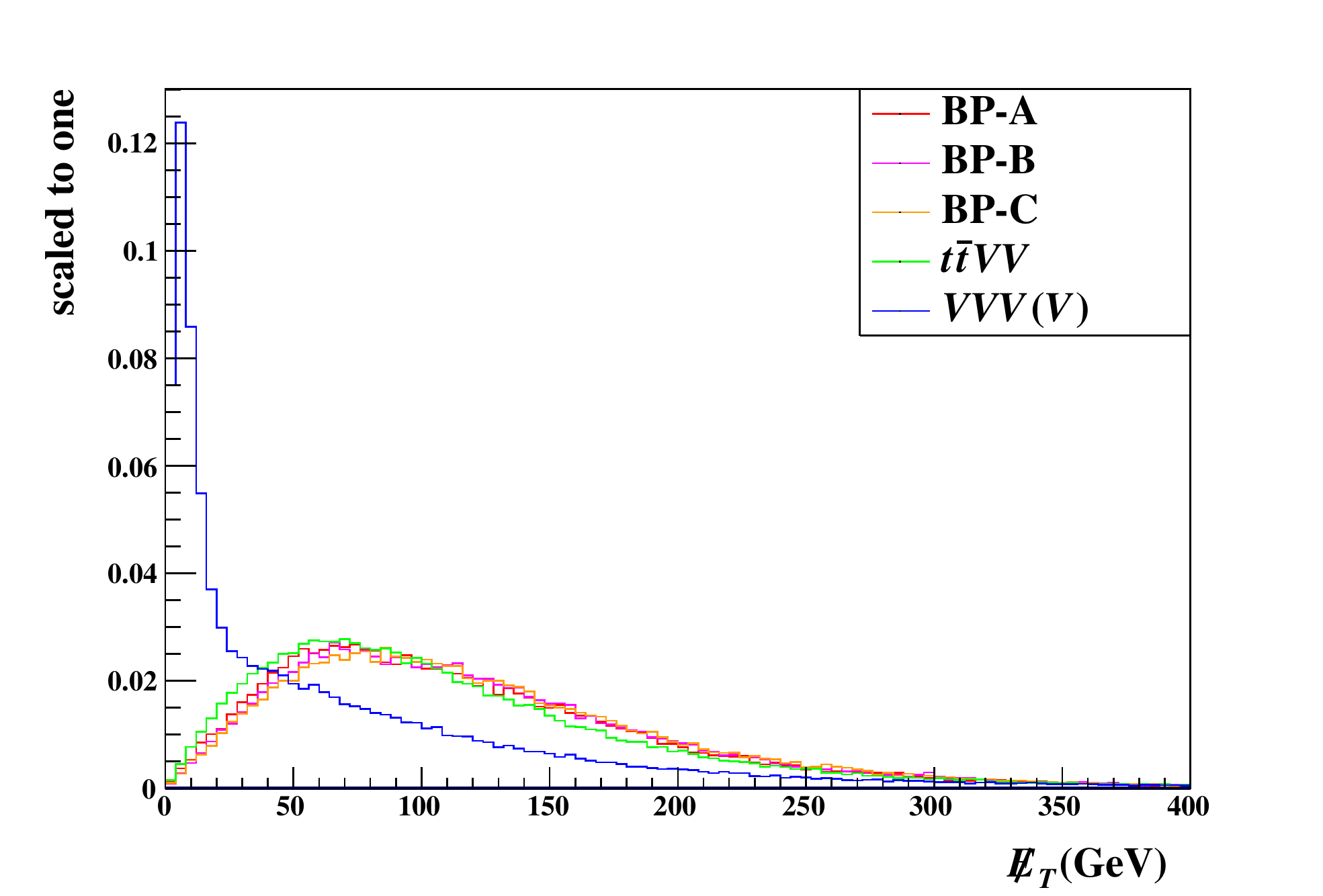}
    \includegraphics[scale=0.4]{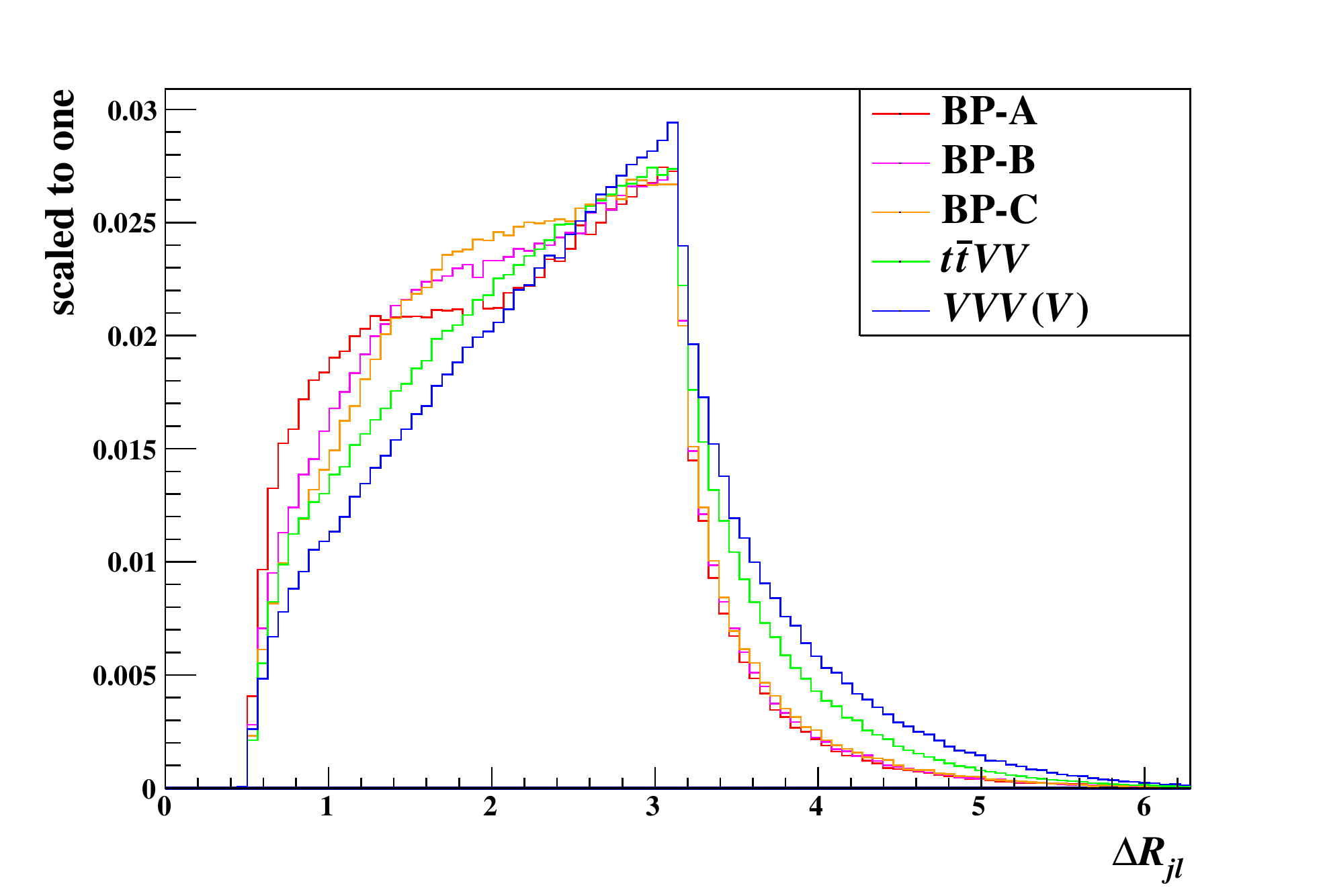}
\caption{Distributions of numbers of same-sign leptons $N(\ell^\pm)$ and jets $N(j)$, transverse momenta $p_T(j)$ and $p_T(\ell)$, missing transverse energy $\cancel{E}_T$, and relative distances $\Delta R_{j\ell}$ for the SST signature and corresponding backgrounds at $13~\TeV$ LHC.}
\label{fig:tri_distributions_13}
\end{figure}

The SST signature, behaving as $3\ell^\pm 4j+\cancel{E}_T$, is the most distinct one in this $\nu$2HDM. It originates from the associated production $H^\pm H$ and $H^\pm A$:
\begin{equation}
p p \to H^\pm H/A \to \ell^\pm N_R \nu N_R\to \ell^\pm \ell^\pm W^\mp \nu \ell^\pm W^\mp
\to 3\ell^\pm 4j + \cancel{E}_T.
\end{equation}
Such a kind of signature is rather exotic in the SM as it violates the lepton number notably by three units in visible final states. The dominant backgrounds are generated from $t\bar{t}W^\pm Z$ and $t\bar{t}ZZ$, where the $t\bar{t}$ pair decays semi-leptonically and the bosons decay leptonically with one lepton missed from each $Z$. The cross sections of backgrounds are extremely small, making the SST signature easier to be tested on LHC.

In Fig.~\ref{fig:tri_distributions_13} we present the distributions of the numbers of same-sign leptons $N(\ell^\pm)$ and jets $N(j)$, transverse momenta $p_T(j)$ and $p_T(\ell)$, missing transverse energy $\cancel{E}_T$, and relative distances $\Delta R_{j\ell}$ for the SST signature at $13~\TeV$ LHC. The results are similar at $14~\TeV$. We impose the same basic cuts in Eq.~(\ref{eqn:basic cuts}) as for the SSD signature. In order to separate out the signal from background, we might apply cuts such as $N(\ell^\pm)=3,~N(j)=4$ and $\cancel{E}_T>30~\GeV$ to isolate the desired signature. But as mentioned above, since the background is rather clean while the signal itself has also a relatively small cross section, only few events could survive the exact selection of $3\ell^\pm 4j+\cancel{E}_T$. We thus turn to a looser selection to pick up inclusive events,
\begin{equation}
\label{eq:sc_tri}
	 N(\ell^\pm)=3,~N(j)\ge 2.
\end{equation}
According to Fig.~\ref{fig:tri_distributions_13} about one third of the signal events could pass this cut. Because of this looser selection, the backgrounds from $VVV$ and $VVVV$ ($V=W/Z$) should be taken into account as well.

    \begin{table}

		\begin{center}
			\begin{tabular}{|c|c|c|c|c|c|}
				\hline
				\multicolumn{2}{|c|}{Channels}   &    Basic cuts in Eq: \ref{eqn:basic cuts}&
				$N(\ell^\pm)=3$   & $N(j)\ge 2$ & $S/\sqrt{S+B}$\\
				\hline
				\multirow{2}{*} {BP-A}
				 & NH &     11 (378)& 3.7 (127) & 3.7 (126) & 1.85 (10.8) \\
				 & IH  &    13 (458)& 4.5 (153)  & 4.5 (152) & 2.05 (11.9) \\
				\hline
				\multirow{2}{*}{BP-B}
				&NH    &     2.9 (102)& 1.1 (37) & 1.0 (36) & 0.87 (5.25) \\
				&IH    &     3.5 (124)& 1.3 (44) & 1.3 (44) & 1.02 (5.93) \\
				\hline
				\multirow{2}{*}{BP-C}
				&NH    &     0.96 (36)& 0.38 (13)& 0.37 (13) & 0.45 (2.65)\\
				&IH     &     1.2 (43)  & 0.46 (16)& 0.44 (16) & 0.51 (3.08) \\
				\hline
				\multicolumn{2}{|c|}{$t\bar{t}VV$}&   1 (36)& 0.25 (8.5)& 0.24 (8.3) &--\\
				\multicolumn{2}{|c|}{$VVV(V)$}&   0.5 (16)& 0.15 (5) & 0.08 (2.7) &--\\
				\hline
			\end{tabular}
		\end{center}
		\caption{Same as Table~\ref{tab:dilepton_cuts}, but for the SST signature.}
\label{tab:trilepton_cuts}
	\end{table}

\begin{figure}
	\centering
	\includegraphics[scale=0.4]{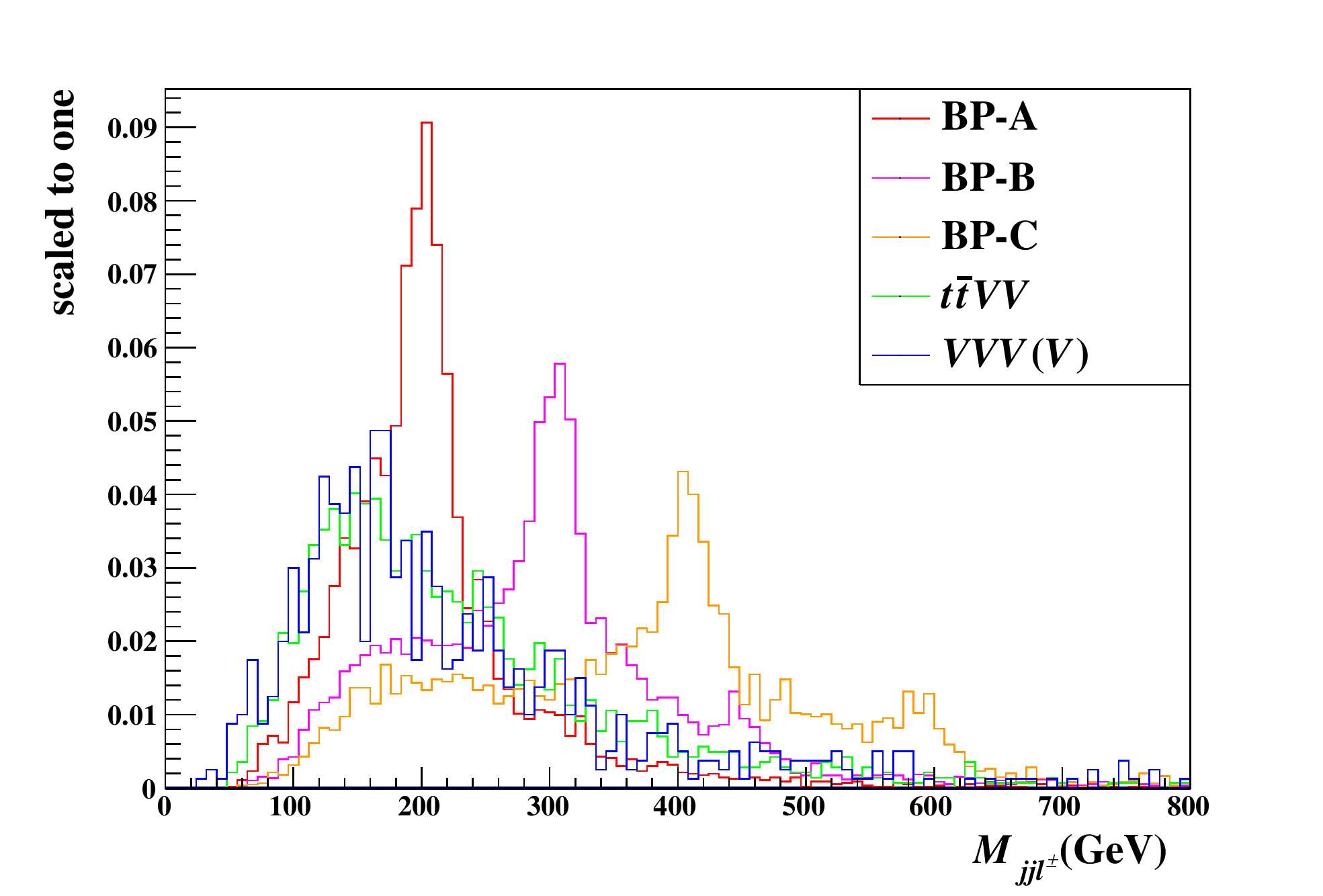}
	\includegraphics[scale=0.4]{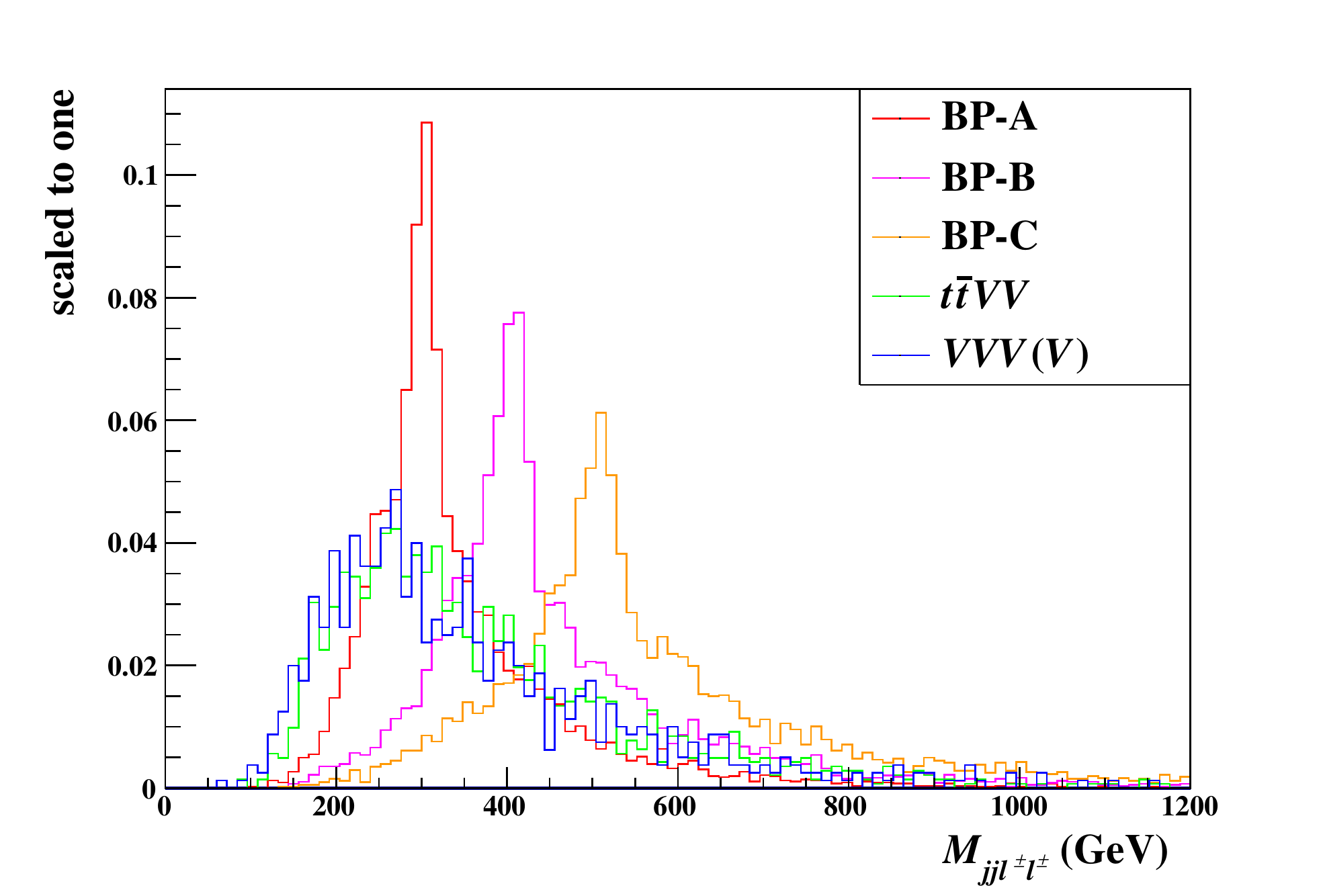}
	\caption{Same as Fig.~\ref{fig:RC_13_di}, but for the SST signature. }
	\label{fig:RC_13_tri}
\end{figure}
	
The cut-flow for the signal at the three benchmark points in Eq.~(\ref{BP}) in both NH and IH cases and the dominant backgrounds is presented in Table~\ref{tab:trilepton_cuts}. The small backgrounds $t\bar{t}W^\pm Z$ and $t\bar{t}ZZ$ are summed to $t\bar{t}VV$, so are $VVV$ and $VVVV$ to $VVV(V)$. At LHC13@100, only BP-A in the IH case could lead to a $2\sigma$ excess. But at LHC14@3000, we will have a good chance to discover BP-A and BP-B, and even BP-C will result in about $2.5\sigma$ ($3\sigma$) excess for NH (IH). In Fig.~\ref{fig:RC_13_tri}, we plot the distributions in the invariant masses  $M_{jj\ell^\pm}$ and $M_{jj\ell^\pm\ell^\pm}$ for the reconstruction of the $N_R$ and $H^\pm$ particles at $13~\TeV$ LHC.

\begin{figure}
		\centering
		\includegraphics[width=0.45\linewidth]{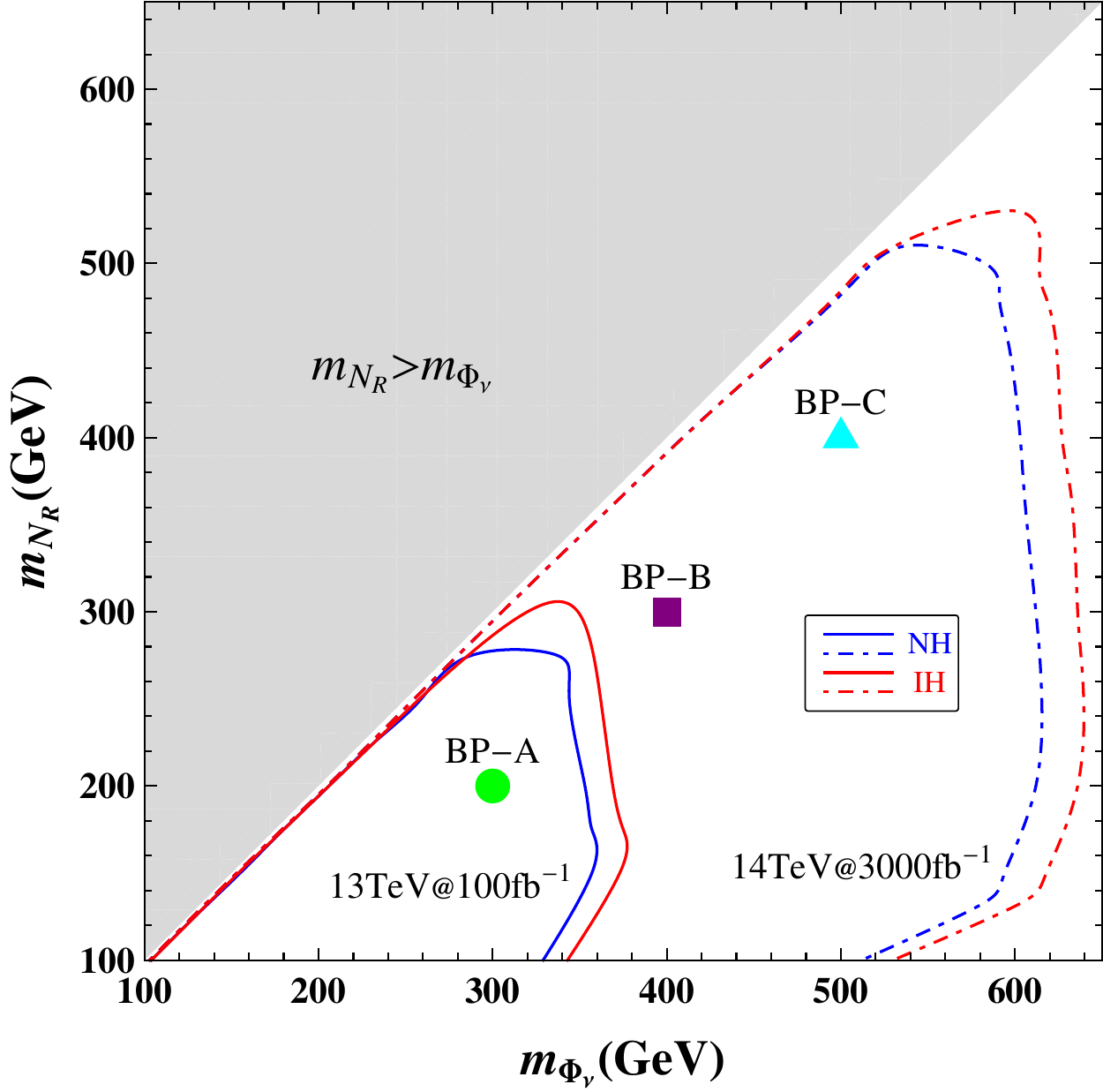}
		\caption{Same as Fig.~\ref{fig:dilepton_exclusion}, but for the SST signature.}
		\label{fig:excludingline}
\end{figure}

Based on the cuts in Eqs.~(\ref{eqn:basic cuts},\ref{eq:sc_tri}) we now acquire the expected 95\% C.L. exclusion limits at LHC13@100 and LHC14@300 by scanning in the $m_{N_R}-m_{\Phi_\nu}$ plane. Our results are presented in Fig.~\ref{fig:excludingline}. For instance, at LHC13@100, $m_{\Phi_\nu}$ could be excluded up to about $350~\GeV$ for NH or $380~\GeV$ for IH at $m_{N_{R}}\sim 150~\GeV$, or conversely, $m_{N_{R}}$ could be excluded up to about $280~\GeV$ for NH or $305~\GeV$ for IH at $m_{\Phi_\nu}\sim 330~\GeV$. With higher integrated luminosity at LHC14@3000, the exclusion limit would extend to $m_{\Phi_\nu}\lesssim 600~\GeV$, $m_{N_R}\lesssim 500~\GeV$ for NH, and even higher for IH, i.e., $m_{\Phi_\nu}\lesssim 640~\GeV$, $m_{N_R}\lesssim 530~\GeV$. It is evident that LHC13@100 (LHC14@3000) will be capable of excluding BP-A (BP-B and BP-C) through the SST signature.

\subsection{Four-lepton Signature}\label{subsec:four-lepton}

Finally we study the most exotic signature involving four leptons, as a result of the decay chains:
\begin{eqnarray}
\label{eq:fourleps}
&&pp\rightarrow H^{+} H^{-}
\rightarrow \ell^{+} N_R \ell^{-} N_R
\rightarrow \ell^{+} W^{\mp} \ell^{\pm} \ell^{-} W^{\mp} \ell^{\pm}
\rightarrow 3\ell^\pm\ell^\mp 4j. 
\end{eqnarray}
This is also the only case that involves two charged scalars in the intermediate state. The major sources of background are $t\bar{t}h,~t\bar{t}V,~t\bar{t}t\bar{t},~t\bar{t}VV$, with $V=W^\pm,~Z$. We found that the kinematical distributions for the four-lepton signature are similar to those for the SST signature. We impose the same basic cuts in Eq.~(\ref{eqn:basic cuts}) but of course drop the cut on $\cancel{E}_T$, because no invisible particles are involved now. In addition, we apply the specific cuts for the four-lepton signature:
\begin{equation}\label{eq:sc_te}
N(j)=4,~N(\ell^{\pm})=3,~N(\ell^{\mp})=1.
\end{equation}
No further cuts are required, since we are practically background free at this stage as will be shown below.

\begin{table}
	\begin{center}
		\begin{tabular}{|c|c|c|c|c|c|}
			\hline
			\multicolumn{2}{|c|}{Channels}    & Basic cuts &
			$N(j)=4$   & $\makecell{N(\ell^\pm)=3 \\ N(\ell^\mp)=1}$\ &
			$S/\sqrt{S+B}$\\
			\hline
			\multirow{2}{*} {BP-A}
			&NH  &  23 (805)    &  5.6 (195)&  0.74 (29)& 0.86 (5.39) \\
			& IH  &    40 (1404)&  9.8 (340)&  1.3 (43)& 1.14 (6.56) \\
			\hline
			\multirow{2}{*}{BP-B}
			&NH    &    7.2 (242)&  1.7 (58)&  0.27 (8.7)& 0.52 (2.95)\\
			&IH    &    13 (422)&  3.0 (102)&  0.47 (15)& 0.69 (3.87)\\
			\hline
			\multirow{2}{*}{BP-C}
			&NH    &    2.4 (89)&  0.58 (21)&  0.09 (3.2)& 0.30 (1.79)\\
			&IH    &    4.2 (155)&  1.0 (37)&  0.16 (5.6)& 0.40 (2.37)\\
			\hline
			\multicolumn{2}{|c|}{$t\bar{t}h$}&   403 (33431)& 36 (2869)&  0 (0)& --\\
			\multicolumn{2}{|c|}{$t\bar{t}t\bar{t}$}&   125 (11316)& 11 (971) &  0 (0) &--\\
			\multicolumn{2}{|c|}{$t\bar{t}V$ }   &    6043 (488674) &  538 (41943)  &  0 (0) &--\\
			\hline
		\end{tabular}
	\end{center}
\caption{Same as Table~\ref{tab:dilepton_cuts}, but for the four-lepton signature.}
\label{tab:fourleps_cuts}
\end{table}

\begin{figure}
	\centering
	\includegraphics[scale=0.4]{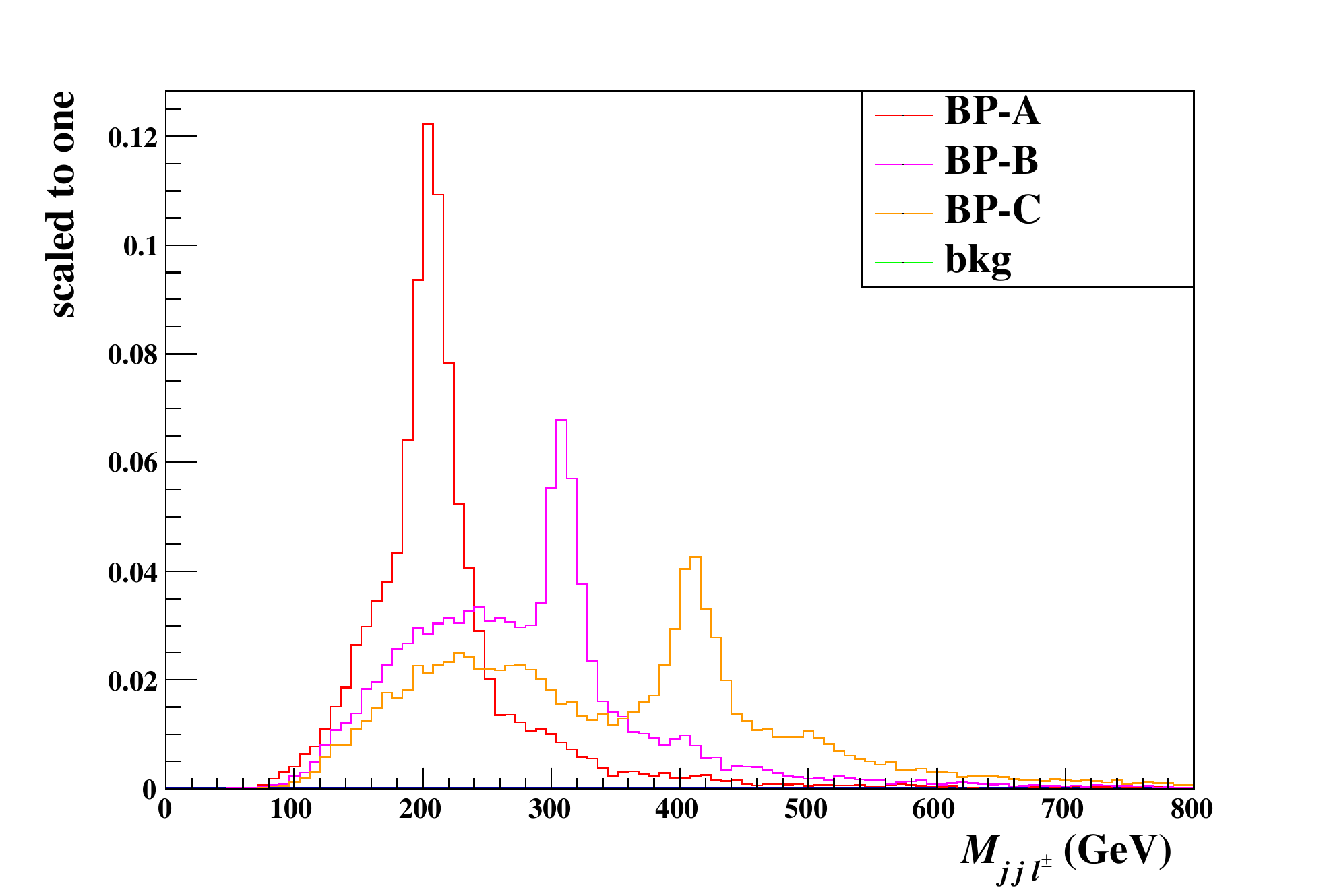}
	\includegraphics[scale=0.4]{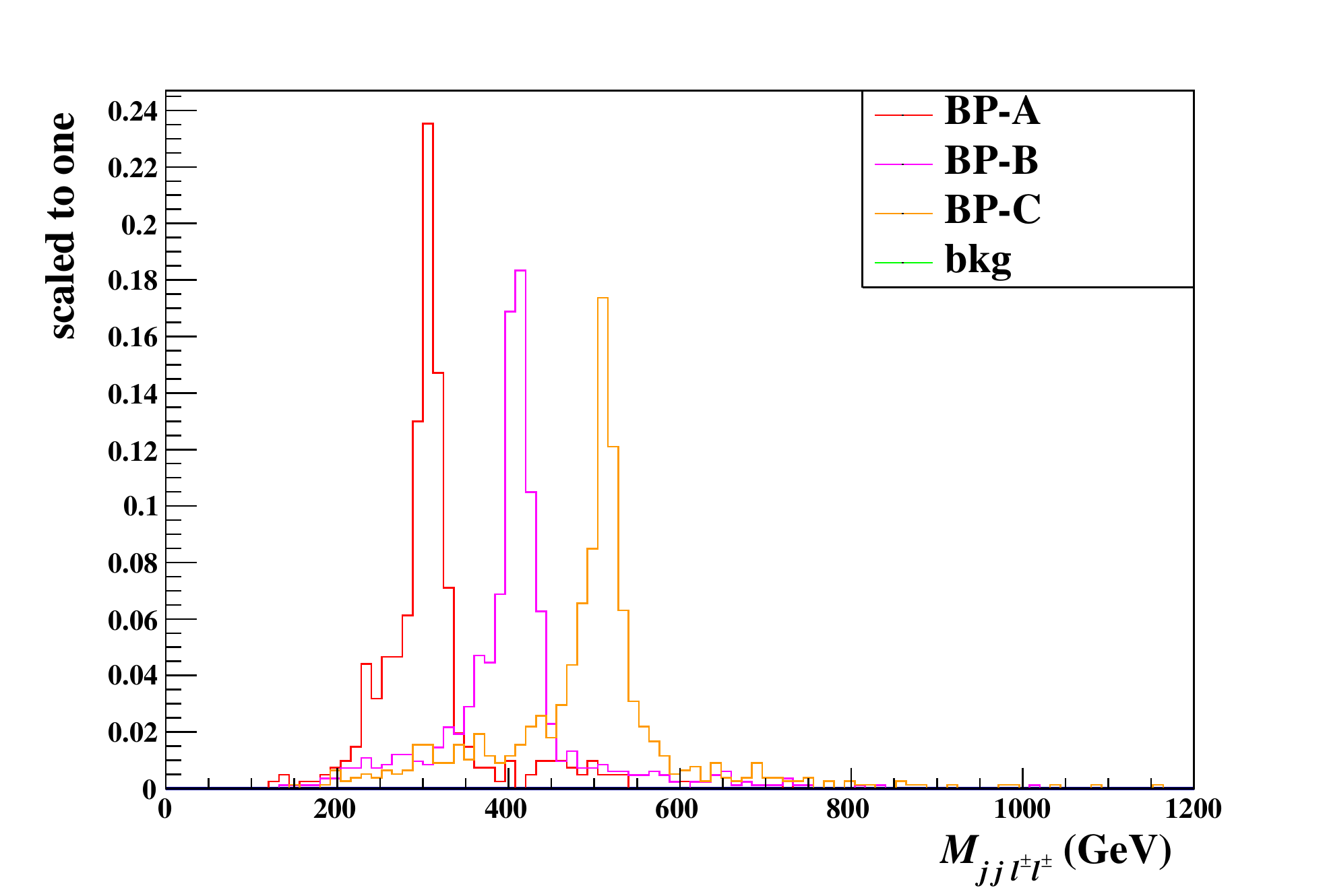}
\caption{Same as Fig.~\ref{fig:dilepton_exclusion}, but for the four-lepton signature.}
\label{fig:RC_13_tetra}
\end{figure}

In Table~\ref{tab:fourleps_cuts} we show the cut-flow for the four-lepton signature at the three benchmark points in Eq.~(\ref{BP}) as well as for the dominant backgrounds. We notice that the backgrounds become zero after the selection cuts in Eqs.~(\ref{eqn:basic cuts},\ref{eq:sc_te}), and this increases the feasibility of the four-lepton signature remarkably. However, we should not forget that this signal is also the weakest among the three studied in this work. At LHC13@100, this signature is undetectable even for BP-A. At LHC14@3000, BP-A will have a $5.39~(6.56)\sigma$ significance in the case of NH (IH), while BP-B can only lead to a $2.95~(3.87)\sigma$ significance. To speak roughly about the prospect to observe such signatures at LHC, the four-lepton signature lies in between the dilepton and trilepton signatures. Adopting the same method as for the other two signatures, we reconstruct in Fig.~\ref{fig:RC_13_tetra} the heavy Majorana neutrinos $N_R$ and charged scalars $H^{\pm}$ via the distributions in the invariant masses $M_{jjl^{\pm}}$ and $M_{jjl^{\pm} l^{\pm}}$ respectively. Although the peaks in the normalized distributions are most remarkable among the three, the number of signal events is also the least.

\begin{figure}
	\begin{center}
		\includegraphics[width=0.45\linewidth]{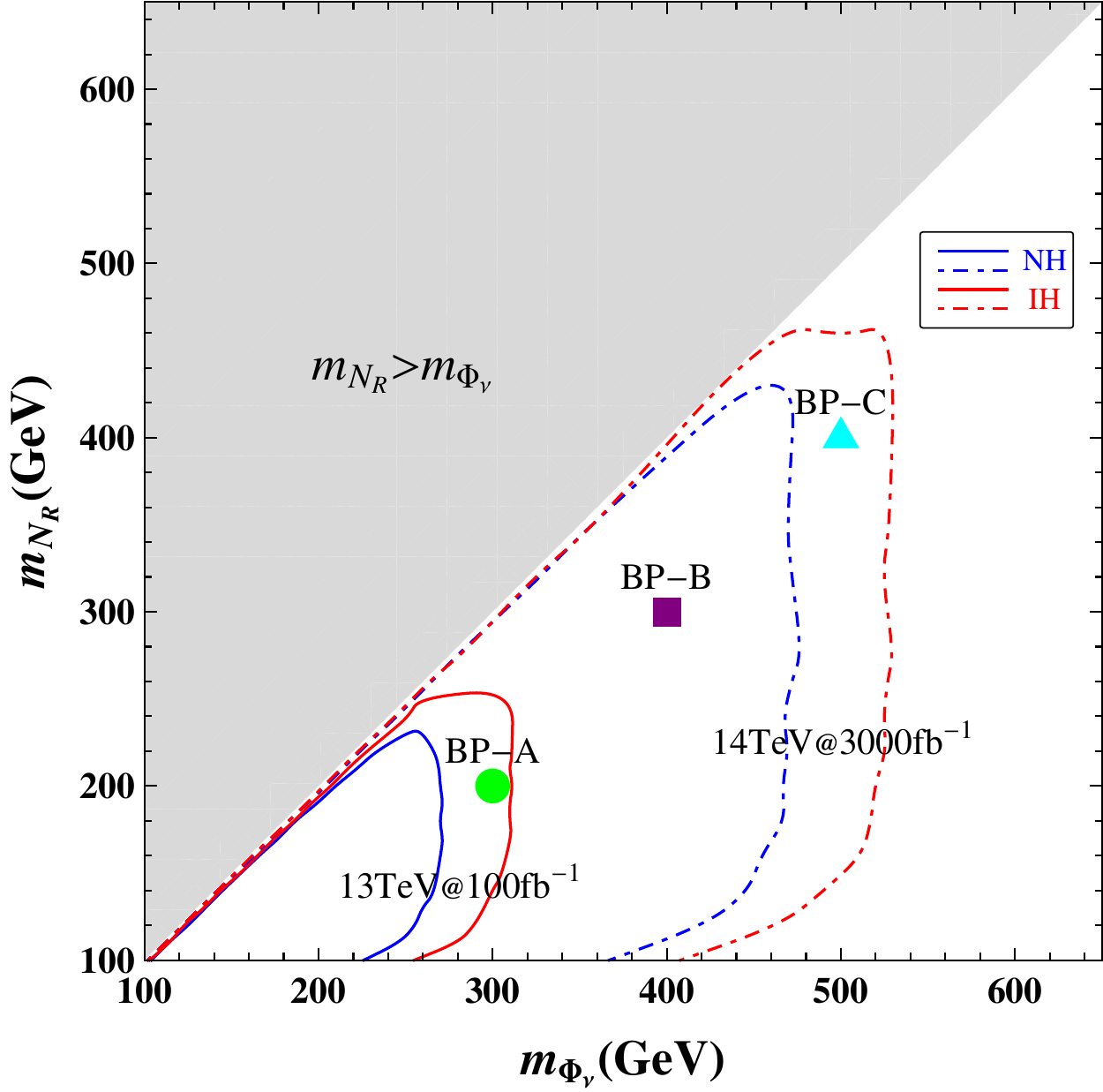}
	\end{center}
\caption{Same as Fig.~\ref{fig:dilepton_exclusion}, but for the four-lepton signature.}
\label{fig:exclusion-tetra}
\end{figure}

The expected 95\% C.L. exclusion limits for the four-lepton signature are shown in Fig.~\ref{fig:exclusion-tetra}. The region with $m_{N_R}\lesssim 250~\GeV$ and $m_{\Phi_\nu}\lesssim 300~\GeV$ could be excluded at LHC13@100, while LHC14@3000 could reach $m_{N_R}\lesssim 450~\GeV$ and $m_{\Phi_\nu}\lesssim 500~\GeV$. Interestingly, we find that BP-A for IH but not for NH could be excluded by LHC13@100, while BP-B would be fully covered by LHC14@3000 for both NH and IH cases.

\section{Conclusion}\label{CL}

The neutrinophilic two-Higgs-doublet model provides a natural way to generate tiny neutrino mass with TeV scale Majorana neutrinos $N_R$ and scalars $\Phi_\nu$, and results in interesting phenomena such as LFV processes and LNV signatures at LHC. We found that the LFV processes of the charged leptons mediated by the Yukawa coupling $\bar{L}\widetilde{\Phi}_\nu N_{R}$ can be within the reach of current experiments. Using the most stringent constraint on the decay $\mu \to e \gamma$, we derived a combined tight bound on the $\Phi_\nu$ mass and VEV, $m_{H^+} v_\nu\gtrsim 600~\GeV\cdot\MeV$, for $N_R$ of a few hundreds GeV.

Our work has been focused on the LNV signatures from heavy Majorana neutrinos at LHC. To achieve this, we systematically investigated the decay properties of the neutrinophilic scalars $H^\pm$, $H$, $A$ and heavy Majorana neutrinos $N_R$ for the mass order $m_{N_R}<m_{\Phi_\nu}$. Our results show that there is strong correlation between the neutrino mass hierarchy and the flavor fraction of charged leptons in the decays of $H^\pm$ and $N_R$. In particular, we expect $\mu^\pm\mu^\pm$ ($e^\pm e^\pm$) dominance for NH (IH) in the LNV decay chain $H^\pm\to \ell^\pm N_R \to \ell^\pm\ell^\pm jj$ $(\ell=e,\mu)$. In addition, the production rate for LNV signatures is larger in the case of IH than NH, making the physics signals in the former case more promising to be detected at LHC.

When the new scalars are heavier than the new neutrinos, they are first produced at LHC via the Drell-Yan processes and then cascade decay into the new neutrinos $H^\pm\to\ell^\pm N_R$, $H/A\to \nu N_R$ and the SM particles $N_R\to\ell^\pm W^\mp$, $\nu Z$, $\nu h$. This results in three kinds of LNV signatures: $2\ell^\pm4j+\cancel{E}_T$, $3\ell^\pm4j+\cancel{E}_T$, and $3\ell^\pm\ell^\mp4j$. To illustrate the testability of such LNV signatures at LHC, we worked with three benchmark points (BP-A, -B, -C, for short), $(m_{N_R},m_{\Phi_\nu})=(200,300),~(300,400),~(400,500)~\GeV$, and performed detailed simulations. We found that the SST signature is the most promising among the three. Although at LHC13@100 it is hard to observe excess events of the SST signature even for BP-A, we can discover BP-A and BP-B from the SST signature at LHC14@3000. Furthermore, BP-C with both scalars and neutrinos much heavier can also lead to about $2.5\sigma$ ($3\sigma$) excess for NH (IH) at LHC14@3000. With such high integrated luminosity there is also a good chance to probe the SSD and four-lepton signatures for BP-A. Conversely, if no excess is observed, the mass region of $m_{N_R}\lesssim 300~\GeV$ and $m_{\Phi_\nu}\lesssim 350~\GeV$ will be excluded by LHC13@100, and the excluded region will be extended at LHC14@3000 to $m_{N_R}\lesssim 500~\GeV$ and $m_{\Phi_\nu}\lesssim 600~\GeV$ by the SST signature.

\section*{Acknowledgement}

This work was supported in part by the Grants No. NSFC-11025525, No. NSFC-11575089 and by the CAS Center for Excellence in Particle Physics (CCEPP). Part of numerical analysis was done with the HPC Cluster of SKLTP/ITP-CAS.


\end{document}